%% file: main.tex
\documentclass{article}

\usepackage{microtype}
\usepackage{fontawesome5}
\usepackage{xcolor}
\usepackage{makecell}
\usepackage{graphicx}
\usepackage{subcaption}
\usepackage{pifont}
\usepackage{tcolorbox}
\usepackage{listings}
\usepackage{wrapfig}
\usepackage{booktabs} %
\usepackage{xspace}
\usepackage{float}
\usepackage{tabularx}
\usepackage{listings}
\usepackage{hyperref}
\usepackage{booktabs}
\usepackage{array}

\newcommand{\methodname}{\textsc{CodeScout}\xspace}
\usepackage{booktabs}
\usepackage{multirow}
\usepackage[table]{xcolor}
\definecolor{highlightcolor}{gray}{0.92}
\definecolor{lightgolden}{rgb}{1.0, 0.95, 0.85} %
\definecolor{headergray}{rgb}{0.92, 0.92, 0.92} %
\definecolor{headerblue}{rgb}{0.88, 0.95, 1.0}  %

\usepackage[preprint]{icml2026}

\usepackage{amsmath}
\usepackage{amssymb}
\usepackage{mathtools}
\usepackage{amsthm}

\usepackage[capitalize,noabbrev]{cleveref}

\theoremstyle{plain}

\theoremstyle{definition}

\theoremstyle{remark}

\usepackage[textsize=tiny]{todonotes}
\usepackage{multirow}
\usepackage{color-edits}
\addauthor{ls}{purple}
\addauthor{soni}{orange}
\addauthor{ty}{teal}
\addauthor{yc}{cyan}
\addauthor{gn}{magenta}

\icmltitlerunning{\methodname: An Effective Recipe for Reinforcement Learning of Code Search Agents}

\begin{document}

\twocolumn[
  \icmltitle{\methodname: \\
  An Effective Recipe for Reinforcement Learning of Code Search Agents
  }

  \icmlsetsymbol{equal}{*}
  \begin{icmlauthorlist}
    \icmlauthor{Lintang Sutawika}{equal,cmu}
    \icmlauthor{Aditya Bharat Soni}{equal,cmu}
    \icmlauthor{Bharath Sriraam R R}{ind}
    \icmlauthor{Apurva Gandhi}{cmu}
    \icmlauthor{Taha Yassine}{ind}
    \icmlauthor{Sanidhya Vijayvargiya}{cmu}
    \icmlauthor{Yuchen Li}{ind}
    \icmlauthor{Xuhui Zhou}{cmu}
    \icmlauthor{Yilin Zhang}{cmu}
    \icmlauthor{Leander Melroy Maben}{cmu}
    \icmlauthor{Graham Neubig}{cmu,openhands}
  \end{icmlauthorlist}
  \centering \faGithub\ \textbf{Code:} \url{https://github.com/OpenHands/codescout}

$\vcenter{\hbox{\resizebox{!}{0.5cm}{\includegraphics{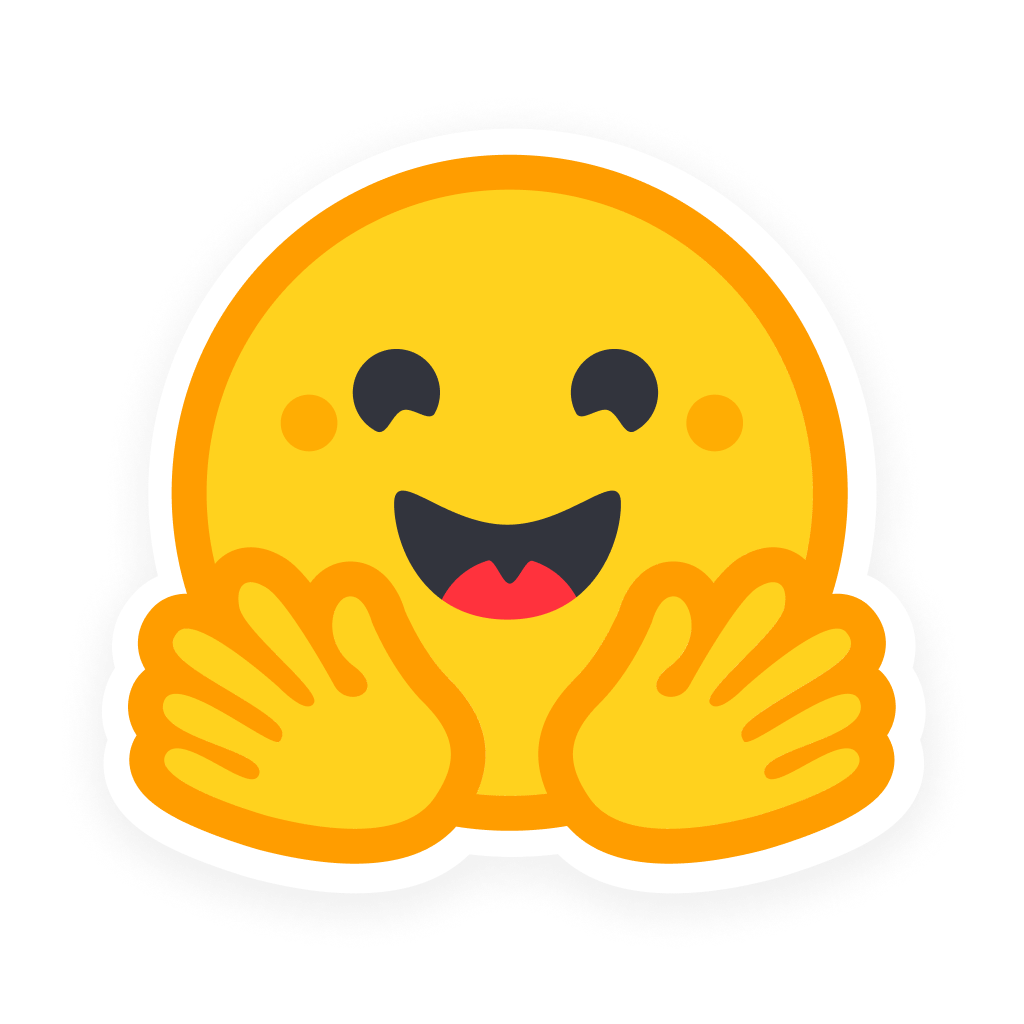}}}}$ 
\textbf{Models \& Data:} \url{https://huggingface.co/collections/OpenHands/codescout}
  \icmlaffiliation{cmu}{Carnegie Mellon University}
  \icmlaffiliation{openhands}{OpenHands}
  \icmlaffiliation{ind}{Independent}

  \icmlcorrespondingauthor{}{lsutawik, adityabs, gneubig}

  \icmlkeywords{Machine Learning, ICML}

  \vskip 0.3in
]

\printAffiliationsAndNotice{\icmlEqualContribution}
\input{Sections/abstract}
\input{Sections/1_introduction}

\input{Sections/2_task_background}

\input{Sections/3_method}
\input{Sections/4_experimental_setup}
\input{Sections/5_results}

\input{Sections/6_analysis}

\input{Sections/8_conclusion}
\input{Sections/9_impact_statement}

\bibliography{reference}
\bibliographystyle{icml2026}

\newpage
\input{Sections/appendix}

\end{document}

%% file: Sections/abstract.tex
\begin{abstract}
A prerequisite for coding agents to perform tasks on large repositories is code localization - the identification of relevant files, classes, and functions to work on. 
While repository-level code localization has been performed using embedding-based retrieval approaches such as vector search, recent work has focused on developing agents to localize relevant code either as a standalone precursor to or interleaved with performing actual work.
Most prior methods on agentic code search equip the agent with complex, specialized tools, such as repository graphs derived from static analysis.
In this paper, we demonstrate that, with an effective reinforcement learning recipe, a coding agent equipped with \emph{nothing more} than a standard Unix terminal can be trained to achieve strong results. Our experiments on three benchmarks (SWE-Bench Verified, Pro, and Lite) reveal that our models consistently achieve superior or competitive performance over \textbf{2-18$\times$} larger base and post-trained LLMs and sometimes approach performance provided by closed models like Claude Sonnet, even when using specialized scaffolds.
Our work particularly focuses on techniques for re-purposing existing coding agent environments for code search, reward design, and RL optimization. We release the resulting model family, \methodname, along with all our code and data for the community to build upon.
\end{abstract}

%% file: Sections/1_introduction.tex
\section{Introduction}

\input{Figures/main_figure}
For repository-level coding tasks such as those in the popular SWE-Bench benchmark \citep{jimenez2024swebenchlanguagemodelsresolve}, a critical first step is \textbf{code localization}: given an issue description and a code repository, the system must identify the relevant files and finer-grained code entities (e.g., classes and functions) to modify~(\S\ref{sec:background}; \citet{husain2019codesearchnet,xia2024agentless}).
This is challenging in large repositories with complex inter-dependencies, and relying on general-purpose language models to solve this localization problem as part of the agentic coding loop can result in high costs, incorrect fixes, and code bloat~\citep{hong2025context}.

To address this issue, it is also common to incorporate some variety of specialized localization module to acquire the relevant context from large codebases in a more efficient and effective manner.
Traditionally, this code search was performed through semantic code search using vector databases \citep{xia2024agentless,wang2025coderag,xie2025swe}. In contrast, recent methods have investigated \emph{agentic code search} - using agents to iteratively navigate the repository and uncover necessary evidence for solving the issue under consideration.
Typically, these methods have involved significant modifications to the agent itself to incorporate static analysis of the codebase, such as LocAgent's repository graph navigation \citep{chen-etal-2025-locagent} and RepoNavigator's ``jump'' tool that retrieves definitions of Python symbols \citep{zhang2025one}.
While well-grounded, these necessitate the use of static analysis tools tailored to a \emph{particular} programming language, increasing operational complexity of deploying such agents on a broader variety of coding scenarios.
Simultaneously, there have been various anecdotal reports from industry regarding reinforcement learning methods that increase the ability of agents to perform localization rapidly \citep{cognition2025swegrep,cursor2025composer}, with varying levels of detail but without a clear training recipe.

Given this background, we address the fundamental research question: \emph{given an appropriate reinforcement learning algorithm, what is an effective recipe to train a code localization agent that simply uses the terminal tool typical of more generic coding agents, yet achieves competitive or superior accuracy over agents using special-purpose tools?}
In this work, we provide the first public demonstration of such a recipe, which involves scalable methods for data and environment creation (\S\ref{sec:data_creation}), a standard agent scaffold (\S\ref{sec:agent_scaffold}), as well as careful attention to reward design (\S\ref{sec:reward_design}) and training algorithm (\S\ref{sec:train_algo}).
In experiments (\S\ref{sec:experiments}, \autoref{sec:results}), we find that the resulting model \methodname achieves localization performance superior to or competitive with existing methods on a variety of localization datasets derived from SWE-Bench Lite, Verified, and Pro.
Importantly, all models and code are released publicly, building a strong baseline for future development in the community to build upon.

%% file: Figures/main_figure.tex
\begin{figure}[!ht]
    \centering
    \includegraphics[width=\columnwidth]{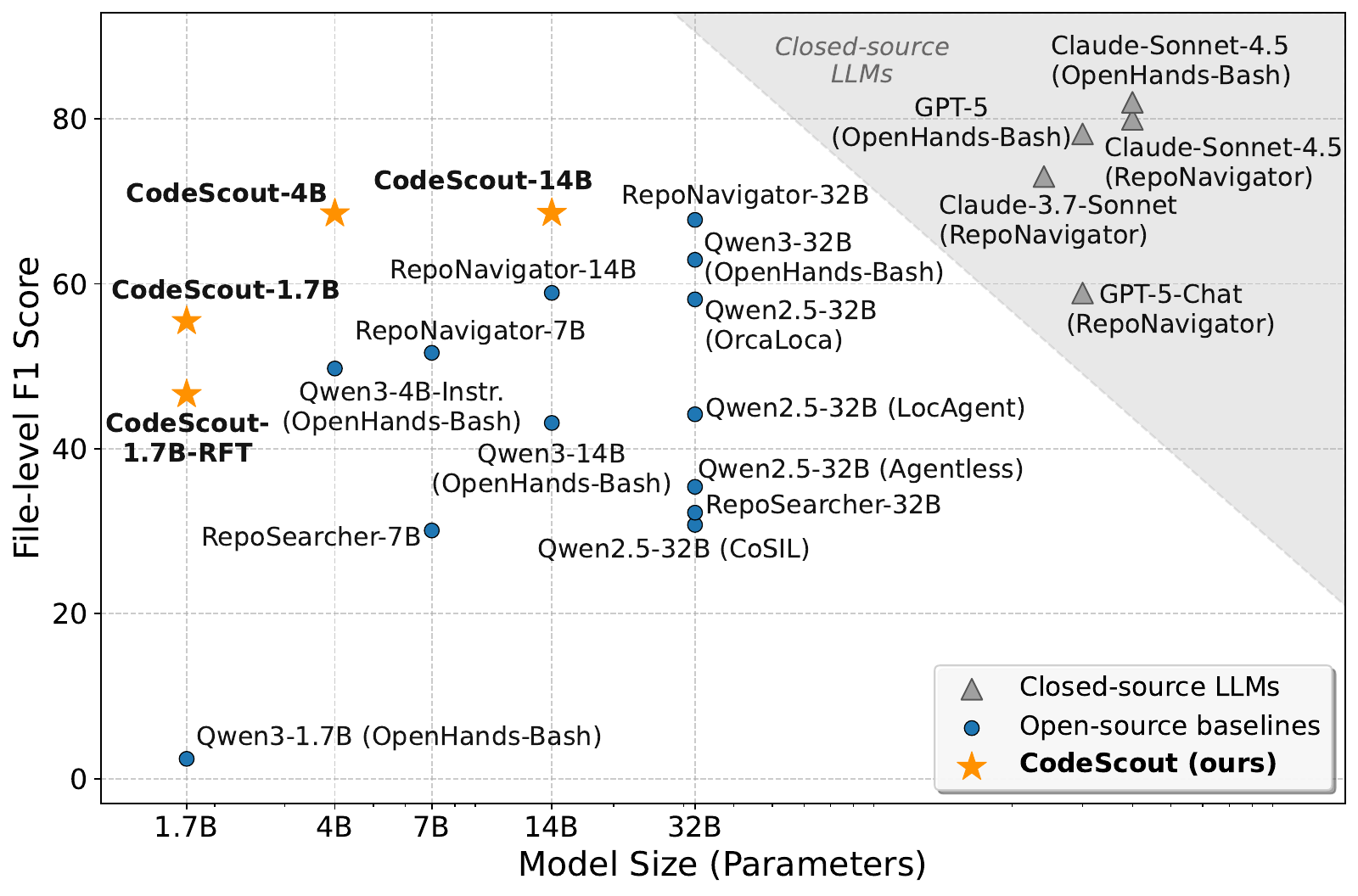}
    \\[3pt]
    \includegraphics[width=\linewidth]{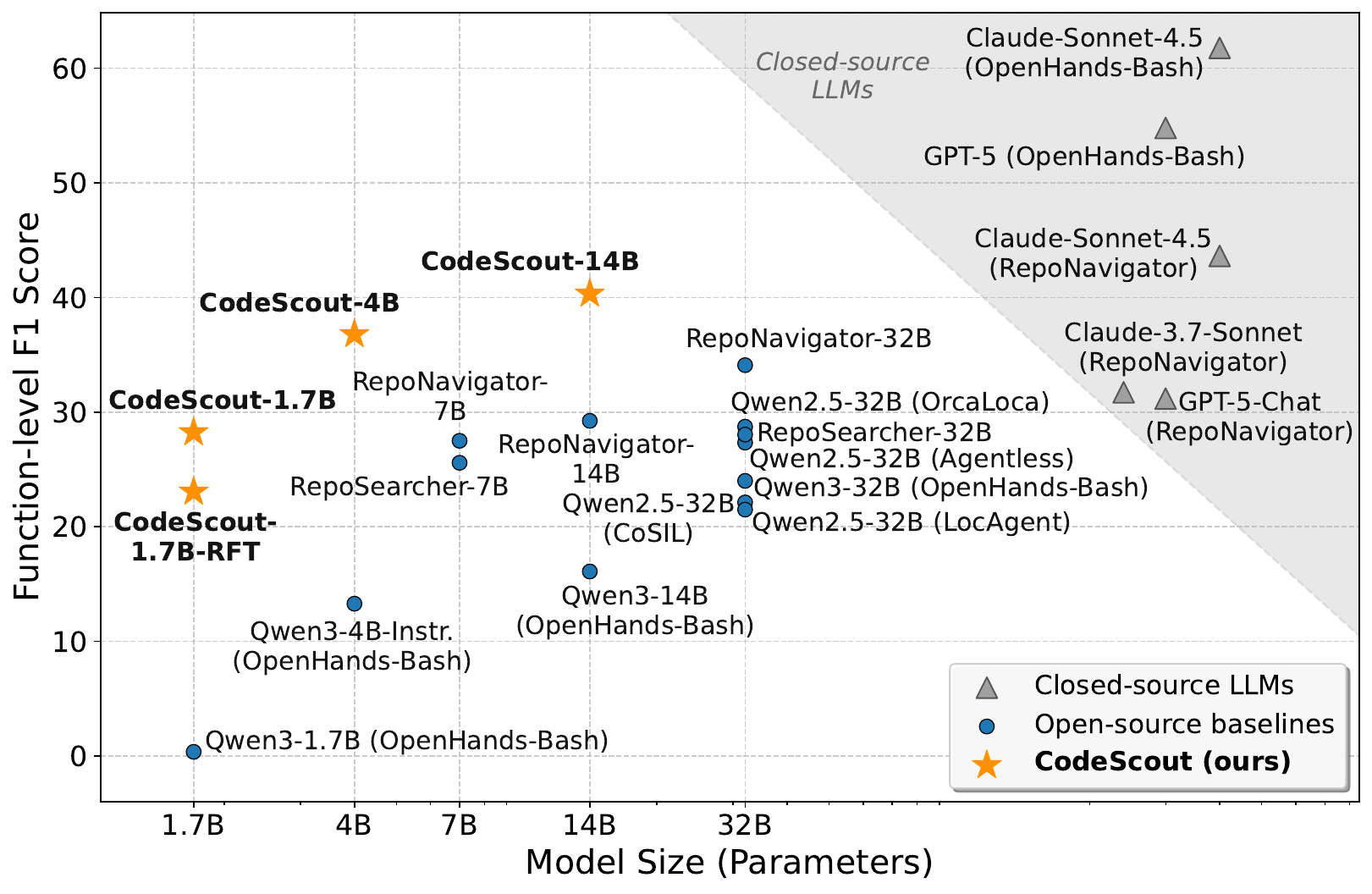}
    \caption{An overview of code localization performance of various approaches on SWE-Bench Verified. \methodname achieves superior or competitive results over larger SoTA open-source LLMs and closes the gap with frontier closed-source LLMs.}
    \label{fig:main}
\end{figure}

%% file: Sections/2_task_background.tex
\input{Tables/baseline_comparison}
\section{Related Work}
\label{sec:background}
Prior work on code localization has proposed specialized agent frameworks and trained open-source LLMs for this task. Table~\ref{tab:baseline_comparison} summarizes prior approaches and compares them with ours. A key limitation of existing methods is that they are restricted to a single programming language, typically Python, due to reliance on language-specific static analysis tools, like AST-based parsers. Extending these approaches to other languages, therefore, requires additional engineering effort. For example, LocAgent~\citep{chen-etal-2025-locagent} and OrcaLoca~\citep{yu2025orcalocallmagentframework} employ language-specific parsers to construct code graphs that capture hierarchical repository dependencies, requiring expensive pre-indexing. CoSIL~\citep{liu2025cosil} dynamically builds module-call graphs to capture ``import'' relationships and function-call graphs to capture ``invoke'' and ``inherit'' relationships. Furthermore, RepoSearcher~\citep{ma2025toolintegratedreinforcementlearningrepo} introduces specialized retrieval tools, for example to extract file imports or search for functions within classes. While CoSIL and RepoSearcher avoid pre-indexing, they still depend on language-specific static analysis to construct call graphs or implement special-purpose tools. Similarly, RepoNavigator~\citep{zhang2025one} provides a ``jump'' tool that resolves Python symbol definitions using a Python language server built using static analysis tools like AST and dependency graphs. In contrast, our approach, \methodname, leverages an agent scaffold that relies \emph{solely} on a standard bash terminal for code localization, which makes it inherently programming language-agnostic by design.

Unlike most prior methods, which either do not train LLMs using their proposed agent frameworks, such as CoSIL and OrcaLoca, or rely on supervised distillation from proprietary closed-source LLMs via rejection sampling fine-tuning~\citep{yuan2023scalingrelationshiplearningmathematical}, like LocAgent and RepoSearcher, \methodname post-trains LLMs directly with reinforcement learning, thereby eliminating dependence on expensive closed-source models for data curation.

Finally, \methodname has a \emph{significantly simpler} agent scaffold both in terms of both engineering overhead and tool count. Since the bash terminal is already a core component of standard coding agents like OpenHands~\citep{wang2025openhandssoftwareagentsdk}, SWE-Agent~\citep{yang2024sweagentagentcomputerinterfacesenable}, and Claude Code~\citep{anthropic_claude_code_2025}, our approach avoids the need to implement specialized task-specific tools or agent scaffolds, unlike prior work. Moreover, when measured by the size of the agent's action space, excluding any exit/finish tools, \methodname uses only 1 tool, compared to 3-5 tools in prior methods. Refer to Appendix~\ref{appendix:related_work_detailed} for a more comprehensive review of prior work.

%% file: Tables/baseline_comparison.tex
\newcommand{\cmark}{\textcolor{green!80!black}{\ding{51}}}
\newcommand{\xmark}{\textcolor{red!80!black}{\ding{55}}}
\begin{table}[t]
\centering
\caption{Comparison of repository-level code localization methods. \methodname is the only approach that directly post-trains LLMs with RL using a simple, programming language-agnostic agent scaffold equipped \emph{solely} with a bash terminal.}
\label{tab:baseline_comparison}
\renewcommand{\arraystretch}{1.2}
\setlength{\tabcolsep}{1.5pt} %
\resizebox{\columnwidth}{!}{%
\begin{tabular}{lccc}
\toprule
\textbf{Method} & 
\textbf{\begin{tabular}[c]{@{}c@{}}Language-Agnostic\\Scaffold\end{tabular}} & 
\textbf{\begin{tabular}[c]{@{}c@{}}Pure RL\\post-training\end{tabular}} & 
\textbf{\# Tools} \\ 
\midrule
LocAgent \citep{chen-etal-2025-locagent}      & \xmark & \xmark & 3 \\
CoSiL \citep{liu2025cosil}                    & \xmark & \xmark & 3 \\
OrcaLoca \citep{yu2025orcalocallmagentframework} & \xmark & \xmark & 5 \\
RepoSearcher \citep{ma2025toolintegratedreinforcementlearningrepo} & \xmark & \xmark & 5 \\
RepoNavigator \citep{zhang2025one}            & \xmark & \cmark & 1 \\
\textbf{\methodname} \textit{\textbf{(Ours)}} & \cmark & \cmark & 1 \\
\bottomrule
\end{tabular}%
}
\end{table}

%% file: Sections/3_method.tex
\input{Figures/system_diagram}
\section{\methodname: An Effective RL Recipe for Code Localization}

This section presents the methodology used to train \methodname, covering training data curation and environment construction (\S\ref{sec:data_creation}), the agent scaffold and tools (\S\ref{sec:agent_scaffold}), reward design (\S\ref{sec:reward_design}), and the overall RL training setup (\S\ref{sec:train_algo}). Figure~\ref{fig:system_diagram} provides an overview of our approach.

\subsection{Data and Environment Curation}
\label{sec:data_creation}
We describe our approach to curating data and constructing RL environments for training \methodname.

Given a GitHub issue $\mathcal{I}$ in a Python repository $\mathcal{R}$, we process the ground-truth issue resolution patch $\mathcal{P}$ to extract the localization targets at three granularities. Specifically, we define the ground-truth target as $y^{\star} = (F^{\star}, M^{\star}, U^{\star})$, where $F^{\star} = {f_1^{\star}, \dots, f_{N_f}^{\star}}$ is the set of modified files, $M^{\star} = {m_1^{\star}, \dots, m_{N_m}^{\star}}$ is the set of modified modules, and $U^{\star} = {u_1^{\star}, \dots, u_{N_u}^{\star}}$ is the set of functions or methods edited by the patch $\mathcal{P}$. An example of these targets is illustrated in Figure~\ref{fig:system_diagram}. The LLM agent is tasked with predicting $y^{\star}$ given the issue description $\mathcal{I}$ and the pre-PR repository state $\mathcal{R}$. We extract ground truth by using patch-processing scripts from LocAgent~\citep{chen-etal-2025-locagent} and enhance them to (i) detect additions of member functions and class attributes at the module and file level, (ii) capture modifications to import statements and global variables at the file level, and (iii) ignore edits to docstrings within functions and classes.

We curate training instances by processing GitHub issues collected by prior work~\citep{yang2025swesmithscalingdatasoftware, pan2025trainingsoftwareengineeringagents}, which were originally intended for training agents to fix issues. We discard issues whose PRs create or delete files, as the agent cannot predict the name of newly created files and we cannot determine ground-truth modules or functions for deleted files. We also ignore non-Python files (e.g., \texttt{README.md}) in the ground truth because function- and module-level information cannot be extracted from them. We also discard instances with empty issue descriptions.

We construct the RL environment by cloning the pre-PR commit of the repository to a specific location known to the agent via its prompt. Since the localization task does not require executing repository code, we do not install project dependencies or use sandboxing/containerization. Notably, since our agent scaffold only requires a terminal, environment setup overhead is significantly lower than scaffolds from prior work whose tools require code graphs, vector databases, or dependency parsers.

\subsection{OpenHands-Bash: Our Agent Scaffold}\label{sec:agent_scaffold}

We use the OpenHands Software Agent SDK~\cite{wang2025openhandssoftwareagentsdk} to implement our scaffold as it performs strongly across software engineering benchmarks, implements core tools (e.g., terminal), supports parallel tool-calling, and offers a modular design that easily integrates with our training backend.

The agent is primarily equipped with a \texttt{Terminal} tool that supports standard Unix commands (e.g., \texttt{rg} (ripgrep), \texttt{find}, \texttt{ls}, \texttt{grep}, \texttt{sed}). We install ripgrep~\citep{ripgrep} in the agent's environment, which is a command-line utility for fast grep-style search using regex patterns recursively through all code files in a directory. In addition, the agent uses a \texttt{LocalizationFinish} tool to submit predicted files, modules, and functions. We refer to this scaffold as \textbf{OpenHands-Bash}.

Our initial experiments used a string-based output format from \citet{chen-etal-2025-locagent}, but we found agents trained with this format to be sensitive to noisy reward signals due to brittle format validation and parsing. We address this by requiring the agent to terminate via the \texttt{LocalizationFinish} tool that enforces a structured output schema and simplifies parsing and validation, improving reward-signal fidelity. Appendix~\ref{appendix:prompts} includes detailed prompts and tool definitions for our agent.

\subsection{Reward Design}\label{sec:reward_design}
For each multi-turn trajectory $\tau$ sampled from the LLM agent, we extract the predicted localization output $y = (F, M, G)$ from the \texttt{LocalizationFinish} tool, which specifies the predicted files, modules, and functions. Given the ground-truth localization $y^{\star} = (F^{\star}, M^{\star}, G^{\star})$, we compute F1 scores at each granularity. Concretely, for each set $S \in \{F, M, G\}$ with the corresponding ground truth $S^{\star} \in \{F^{\star}, M^{\star}, G^{\star}\}$, we compute the F1 score (which is the harmonic mean of precision and recall): $r^{\mathsf{F1-file}}$, $r^{\mathsf{F1-module}}$, and $r^{\mathsf{F1-func}}$. Our reward function is defined as:
\begin{equation}
\begin{aligned}
    r(\tau, y, y^{\star}) &= r^{\mathsf{F1-file}}(y, y^{\star}) + r^{\mathsf{F1-module}}(y, y^{\star}) \\
    &\quad + r^{\mathsf{F1-func}}(y, y^{\star})
\end{aligned}
\label{eqn:reward}
\end{equation}
In early experiments for \methodname-14B, we observed training collapse characterized by near-zero rewards in later stages. Our analysis indicated that the agent frequently exhausted the step budget without submitting predictions. To mitigate this issue, we use an auxiliary binary reward $r^{\mathsf{turn}}(\tau, k)$ that assigns 1 if and only if the agent terminates in exactly $k$ turns, where $k$ is the step limit. This simple mechanism encourages timely termination within the step limit. Accordingly, the final reward for \methodname-14B adds this binary term to $r(\tau, y, y^{\star})$ in Equation~\ref{eqn:reward}. We also explored auxiliary rewards to incentivize parallel tool-calling, motivated by SWE-grep~\citep{cognition2025swegrep}, but found these interventions to hurt overall performance. Thus, we instead explicitly prompt the agent to use parallel tool-calling (Appendix~\ref{appendix:prompts}).

\subsection{RL Training Algorithm}\label{sec:train_algo}

We implement our training backend using SkyRL~\cite{griggs2025skrylv01}, a modular framework for LLM reinforcement learning. SkyRL supports asynchronous training~\citep{fu2025areallargescaleasynchronousreinforcement}, improving GPU utilization by parallelizing rollout generation and weight optimization. Specifically, parameter updates are triggered after collecting a sufficient number of rollouts, allowing trajectories from slightly stale checkpoints. Concretely, trajectories used at the $n^{\text{th}}$ training step may be sampled from checkpoints that are atmost $t$ steps stale, where $t$ is the maximum staleness (set to 4 in our experiments). After each optimization step, we synchronize updated weights with the vLLM~\citep{kwon2023efficient} engines and terminate any in-progress inference requests.

We train \methodname models using Group Sequence Policy Optimization (GSPO)~\citep{zheng2025groupsequencepolicyoptimization}. The loss function is expressed below where $\theta$ denotes model parameters; $i$ indexes sequences in a group of size $G$; $y_i$ is the $i^{th}$ output sequence with length $|y_i|$, and $y_{i,t}$ is its $t^{th}$ token. $\pi_\theta$ and $\pi_{\theta_{\text{old}}}$ are the current and previous policies, $\hat{A}_i$ is the advantage, and $\varepsilon$ is the clipping parameter:
{
\small %
\begin{align}
    \mathcal{J}_{\text{GSPO}}(\theta) &= \nonumber \\ 
    \mathbb{E}_i &\left[ \frac{1}{G} \sum_{i=1}^{G} \min\left( s_i(\theta) \hat{A}_i,\, \text{clip}\left( s_i(\theta), 1-\varepsilon, 1+\varepsilon \right) \hat{A}_i \right) \right]
    \label{eq:gspo}
\end{align}
}
where,  $s_i$ is the importance ratio derived from sequence likelihood~\cite{zheng-etal-2023-click}:
\begin{align}
    s_i(\theta) = \exp\left( \frac{1}{|y_i|} \sum_{t=1}^{|y_i|} \log \frac{\pi_\theta(y_{i,t} | x, y_{i,<t})}{\pi_{\theta_{\text{old}}}(y_{i,t} | x, y_{i,<t})} \right)
    \label{eq:is}
\end{align}
Following DR. GRPO~\cite{liu2025understandingr1zeroliketrainingcritical}, we remove the KL regularization term from the loss and the standard deviation from the advantage calculation.
\begin{align}
    \hat{A}_i = r_i - \text{mean}(\mathbf{r}), \mathbf{r} = \{r_1, ..., r_G\}
\end{align}
We also disable entropy loss and mask loss for rollouts that exhaust maximum steps without calling the finish tool. Appendix~\ref{appendix:rl_ablation_study} includes additional analysis on the effect of the RL algorithm on localization performance.

%% file: Figures/system_diagram.tex
\begin{figure*}[!ht]
    \centering
    \includegraphics[width=0.75\textwidth]{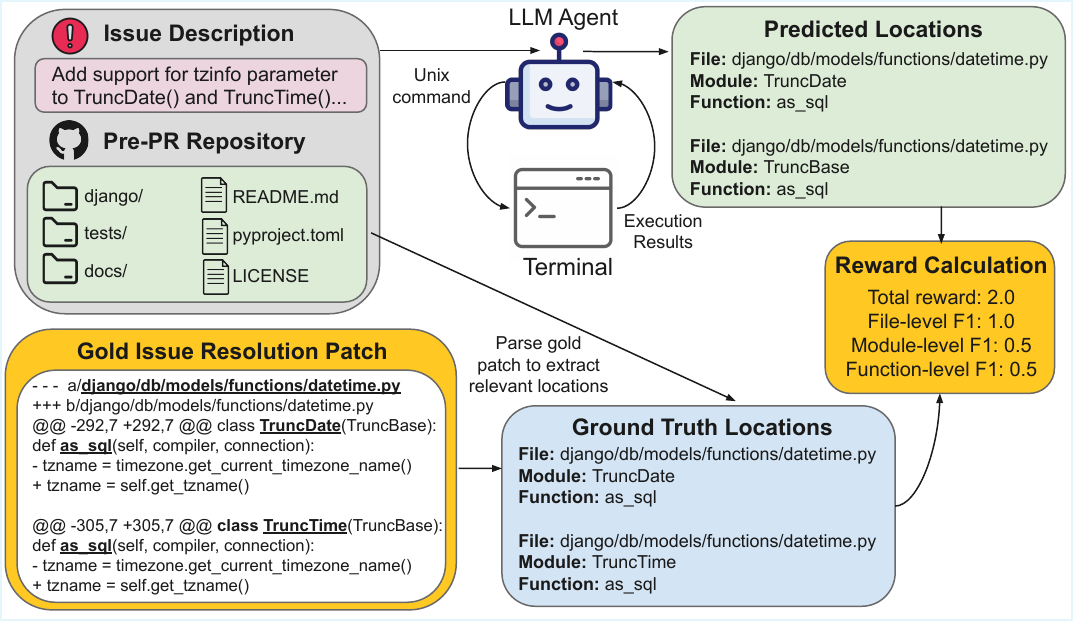}
    \caption{An overview of \methodname: given a GitHub issue, the LLM agent navigates the pre-PR codebase using a terminal and predicts the relevant set of files, modules, and functions. The reward function computes F1 scores for these three granularities using ground truth locations extracted from the gold issue resolution patch.}
    \label{fig:system_diagram}
\end{figure*}

%% file: Sections/4_experimental_setup.tex
\input{Tables/results_summary}

\section{Experimental Setup}
\label{sec:experiments}

This section presents our experimental setup for training (\S\ref{sec:train_setup}) and evaluation (\S\ref{sec:eval_benchmarks}), and describes the baselines we compare \methodname against (\S\ref{sec:baselines}).

\subsection{Training Setup}\label{sec:train_setup}
We process the SWE-Smith~\citep{yang2025swesmithscalingdatasoftware} dataset using the procedure from \S\ref{sec:data_creation}, resulting in a filtered training set of 39K instances across 128 repositories. None of these repositories overlap with those used in our evaluation benchmarks (\S\ref{sec:eval_benchmarks}), avoiding any risk of dataset contamination.

We train models from the Qwen3 family~\citep{yang2025qwen3technicalreport}, specifically \texttt{Qwen3-1.7B}, \texttt{Qwen3-4B-Instruct-2507}, and \texttt{Qwen3-14B}. For the 1.7B and 14B models, we disable thinking and use a modified chat template that preserves the \texttt{<think>} and \texttt{</think>} tokens from earlier turns during tokenization\footnote{https://huggingface.co/OpenPipe/Qwen3-14B-Instruct}, overriding the default behavior that removes them. This modification preserves the sequence extension property where previous trajectory steps are guaranteed to be prefixes for future steps allowing us to merge trajectory steps into a single training sequence, greatly improving training efficiency. We mask the loss for tokens not generated by the model, including the system prompt, user prompt, and environment observations (i.e., tool responses). 

\methodname-4B and \methodname-14B are directly trained from their corresponding base models using our modified GSPO algorithm (\S\ref{sec:train_algo}). For \methodname-4B, we train for 200 steps on 1.6K instances using a batch size of 8 and sample 8 rollouts per instance. The model uses a maximum context length of 40K tokens, the reward function from Equation~\ref{eqn:reward}, and a maximum of 6 turns per episode. For \methodname-14B, we train for 300 steps on 9.6K instances with a batch size of 32, sampling 4 rollouts per instance. We use a maximum context length of 50K tokens, extended with YaRN~\citep{peng2026yarnefficientcontextwindow}, and the reward function from Equation~\ref{eqn:reward} augmented with the auxiliary term $r^{\mathsf{turn}}(\tau, k)$ while $k$ is set to 4 as the agent is allowed a maximum of 4 turns per rollout.

Since the base \texttt{Qwen3-1.7B} model performs very poorly (near-zero F1) on our evaluation benchmarks (\S\ref{sec:results}), we do not directly train it using RL. Instead, we warm-start the model with rejection sampling fine-tuning (RFT)~\citep{yuan2023scalingrelationshiplearningmathematical} on trajectories sampled from \methodname-14B. Specifically, we sample rollouts from \methodname-14B on a random subset of 7.7K training instances and retain only those achieving perfect localization across all three granularities (i.e., F1 = 1.0 for file, module, and function; see \S\ref{sec:eval_benchmarks}), yielding 4K training examples. We then use the veRL framework~\citep{Sheng_2025} to perform supervised fine-tuning of \texttt{Qwen3-1.7B} on these successful trajectories for one epoch, with a learning rate of $5e^{-5}$, a cosine learning rate scheduler, warmup ratio of 0.1, global batch size of 8, and the AdamW optimizer~\citep{loshchilov2019decoupledweightdecayregularization}. The resulting checkpoint (\methodname-1.7B-RFT) is subsequently trained using RL (\S\ref{sec:train_algo}) for 100 steps on 800 instances (not seen during RFT) with a batch size of 8, sampling 8 rollouts per instance. We use a maximum context length of 32K tokens, the reward function from Equation~\ref{eqn:reward}, and allow a maximum of 4 turns per episode.

Across all \methodname RL training runs, we use a constant learning rate of $1e^{-6}$, set \texttt{clip\_ratio\_low} to $3e^{-4}$ and \texttt{clip\_ratio\_high} to $4e^{-4}$, use the AdamW optimizer, and sample rollouts with temperature of 1.0. All experiments are conducted with 8$\times$H100 GPUs. Appendix~\ref{appendix:rl_loss_curves} presents training curves for RL rewards and SFT loss.
\input{Tables/results_verified}

\subsection{Evaluation Setup}\label{sec:eval_benchmarks}

We report performance on three benchmarks: SWE-Bench Verified~\citep{chowdhury2024swebenchverified} (500 instances), SWE-Bench Lite~\citep{jimenez2024swebenchlanguagemodelsresolve} (300 instances), and the Python subset of SWE-Bench Pro~\citep{deng2025swebenchproaiagents} (266 instances), which is substantially more challenging than the other two. For all benchmarks, we re-purpose the datasets and extract ground-truth locations using the method in \S\ref{sec:data_creation}.

During evaluation of Qwen3 models, as well as during rollout generation from \methodname-14B for training \methodname-1.7B-RFT (\S\ref{sec:train_setup}), we use the decoding hyperparameters recommended by the Qwen3 developers. For instruct and reasoning models with thinking disabled, we use temperature $=0.7$, top-$k=20$, and top-$p=0.8$. For reasoning models with thinking enabled, we use temperature $=0.6$, top-$k=20$, and top-$p=0.95$. Finally, we set the maximum context length to 132K tokens for all Qwen3 models.

\paragraph{Evaluation Metrics:} Our primary evaluation metric is the instance-wise average of F1 score between predicted locations and ground-truth locations across three granularities: file, module, and function. In addition, we report average precision and recall for each granularity to compare with baselines that predict a fixed number of locations, which often results in precision-recall disparities.
\subsection{Baselines}\label{sec:baselines}
We compare \methodname against several baselines using both open-source and closed-source LLMs: RepoNavigator~\citep{zhang2025one}, RepoSearcher~\citep{ma2025toolintegratedreinforcementlearningrepo}, LocAgent~\citep{chen-etal-2025-locagent}, OrcaLoca~\citep{yu2025orcalocallmagentframework}, CoSIL~\citep{liu2025cosil}, and Agentless~\citep{xia2024agentless}. All baseline metrics are taken from prior work, either from the corresponding paper or from subsequent work.

Many baselines like ~\citet{ma2025toolintegratedreinforcementlearningrepo,liu2025cosil,chen-etal-2025-locagent,xia2024agentless} output a ranked list of top-$K$ locations (typically $K$ is fixed and set to 5), whereas other approaches like ~\citet{zhang2025one, yu2025orcalocallmagentframework} and \methodname dynamically predicts a variable number of locations, which better aligns with the fact that issues may require edits across varying numbers of locations. As a result, for our benchmarks, some baseline approaches often have greater recall than precision. We argue that our design is preferable because higher precision is more crucial than higher recall for downstream issue resolution - ~\citet{cognition2025swegrep} report that polluting the context of the coding agent is more detrimental than leaving some context out, as the agent is typically only a few searches away from recovering any remaining context. Refer to Appendix \ref{appendix:topk_discussion} for additional discussion.

We also evaluate other LLMs with OpenHands-Bash (\S\ref{sec:agent_scaffold}). First, we benchmark the base variants of \methodname LLMs: \texttt{Qwen3-4B-Instruct-2507}, and the non-thinking versions of \texttt{Qwen3-1.7B} and \texttt{Qwen3-14B}. We also include \texttt{Qwen3-32B} with thinking enabled to compare against a larger reasoning model. Finally, we evaluate two closed-source models, GPT-5 and Claude-Sonnet-4.5. For open-source LLMs, we use the same prompts as \methodname, specifying a turn limit of 4 (or 6 for 4B LLMs) while allowing up to \textbf{15} turns in the backend to ensure a fair comparison, since base models may require more steps to complete the task. For closed-source models, we limit backend turns to 6 due to cost constraints. Despite specifying the turn limit in the system prompt, GPT-5 and Claude Sonnet 4.5 frequently exhaust all steps without submitting predictions, yielding a score of 0. Following RepoNavigator~\citep{zhang2025one}
, we mitigate this by adding a reminder message before the last turn that prompts the LLM to submit its final answer.

%% file: Tables/results_summary.tex
\begin{table*}[t]
\caption{Summary of results on SWE-Bench Verified, Pro, and Lite. For all benchmarks and search granularities, \methodname achieves a new open-source SoTA, with stronger/comparable results over base and post-trained LLMs upto \textbf{8-18$\times$ larger}, narrowing the gap with and often surpassing closed-source LLMs. The highest metric is \textbf{bold-faced} and 2$^{nd}$ highest metric is \underline{underlined}.}
\centering
\begin{tabular}{c|c|cc} 
\toprule
\textbf{Benchmark} & \textbf{Method} & \textbf{File-level F1} & \textbf{Function-level F1} \\
\midrule

& RepoNavigator-7B & 51.63 & 27.49 \\
\rowcolor{lightgolden}\cellcolor{white}& \textbf{\methodname-1.7B} & 55.46 & 28.22 \\
& RepoNavigator-32B & 67.75 & 34.09 \\
\rowcolor{lightgolden}\cellcolor{white}& \textbf{\methodname-4B} & 68.52 & 36.78 \\
\rowcolor{lightgolden}\cellcolor{white}& \textbf{\methodname-14B} & \underline{68.57} & \underline{40.32} \\
\rowcolor{gray!20!white}\cellcolor{white}& RepoNavigator + GPT-5-Chat & 58.88 & 31.17 \\
\rowcolor{gray!20!white}\cellcolor{white}\multirow{-7}{*}{SWE-Bench Verified} & RepoNavigator + Claude-Sonnet-4.5& \textbf{79.94} & \textbf{43.62} \\
\midrule

& RepoNavigator-7B & 39.74 & 14.29 \\
\rowcolor{lightgolden}\cellcolor{white}& \textbf{\methodname-1.7B} & 40.96 & 18.24 \\
& RepoNavigator-32B & \textbf{57.57} & 20.72 \\
\rowcolor{lightgolden}\cellcolor{white}& \textbf{\methodname-4B} & 51.77 & \textbf{29.03} \\
\rowcolor{lightgolden}\cellcolor{white}\multirow{-5}{*}{SWE-Bench Pro} & \textbf{\methodname-14B} & \underline{53.63} & \underline{28.74} \\
\midrule

& OpenHands-Bash + Qwen3-32B (Thinking) & 58.98 & 23.76\\
\rowcolor{lightgolden}\cellcolor{white}&  \textbf{\methodname-1.7B} & 56.57 & 27.07 \\
\rowcolor{lightgolden}\cellcolor{white}& \textbf{\methodname-4B} & \underline{67.03} & \underline{39.87} \\
\rowcolor{lightgolden}\cellcolor{white}\multirow{-4}{*}{SWE-Bench Lite} & \textbf{\methodname-14B} & \textbf{71.84} & \textbf{44.43} \\

\bottomrule
\end{tabular}
\label{tab:swe_bench_summary}
\end{table*}

%% file: Tables/results_verified.tex
\begin{table*}[!t]
\caption{Results on \textbf{SWE-Bench Verified}. \methodname outperforms up to \textbf{8$\times$ larger} base and post-trained LLMs, narrowing the gap with or surpassing closed-source LLMs. Qwen2.5 results use instruct models. $^{rem}$ adds a submission reminder (\S\ref{sec:baselines}), $^\dagger$trained with GRPO/GSPO; $^\ddagger$RFT from \methodname-14B; $^\triangle$RFT from Claude-3.7-Sonnet. The best metric is \textbf{bold-faced} and 2$^{nd}$ best is \underline{underlined}.}
\centering
\renewcommand{\arraystretch}{1.2}
\setlength{\tabcolsep}{4pt}
\resizebox{\textwidth}{!}{%
\begin{tabular}{lc|ccccccccc}
\toprule
\multirow{2}{*}{\textbf{Scaffold}} & \multirow{2}{*}{\textbf{LLM}} & \multicolumn{3}{c}{\textbf{File-level}} & \multicolumn{3}{c}{\textbf{Module-level}} & \multicolumn{3}{c}{\textbf{Function-level}} \\
& & F1 & Prec. & Rec. & F1 & Prec. & Rec. & F1 & Prec. & Rec. \\
\midrule
\rowcolor{headergray}
\multicolumn{11}{c}{\textbf{Closed-Source LLMs}} \\
\midrule
RepoSearcher & Claude-3.7-Sonnet & 32.30 & 20.24 & \textbf{89.24} & - & - & - & 26.91 & 18.64 & \textbf{66.08} \\
\midrule
\multirow{3}{*}{RepoNavigator} & GPT-5-Chat & 58.88 & 61.87 & 58.17 & - & - & - & 31.17 & 34.56 & 30.42 \\
& Claude-3.7-Sonnet & 73.01 & 75.95 & 72.26 & - & - & - & 31.72 & 34.43 & 31.03 \\
& Claude-Sonnet-4.5 & \underline{79.94} & \underline{81.92} & 80.68 & - & - & - & 43.62 & 45.76 &43.97 \\
\midrule
\multirow{2}{*}{OpenHands-Bash} & GPT-5 & 3.20 & 3.20 & 3.20 & 2.60 & 2.60 & 2.60 & 2.60 & 2.60 & 2.60 \\
& Claude-Sonnet-4.5 & 0.80 & 0.80 & 0.80 & 0.40 & 0.40 & 0.40 & 0.40 & 0.40 & 0.40\\
\multirow{2}{*}{OpenHands-Bash$^{rem}$} & GPT-5 & 78.18 & 79.25 & 80.80 & \underline{61.17} & \underline{62.23} & \underline{63.35} & \underline{54.79} & \underline{56.80} & 56.53 \\
& Claude-Sonnet-4.5 & \textbf{82.01} & \textbf{84.50} & \underline{82.86} & \textbf{67.19} & \textbf{70.11} & \textbf{67.47} & \textbf{61.78} & \textbf{65.42} & \underline{61.99} \\
\midrule
\rowcolor{headerblue}
\multicolumn{11}{c}{\textbf{Open-Source LLMs}} \\
\midrule
CoSIL & & 30.77 & 19.34 & \underline{83.50} & - & - & - & 22.11 & 14.85 & 55.38 \\
Agentless &  & 35.38 & 25.60 & 78.93 & - & - & - & 27.33 & 24.07 & 40.97 \\
LocAgent & & 44.18 & 34.18 & 79.39 & - & - & - & 21.48 & 16.29 & 46.79 \\
OrcaLoca &\multirow{-4}{*}{Qwen2.5-32B}  & 58.11 & 59.51 & 59.57 & - & - & - & 28.72 & 25.59 & 39.14 \\
\midrule
\multirow{2}{*}{RepoSearcher} & Qwen2.5-7B$^{\triangle\dagger}$ & 30.09 & 18.80 & 83.11 & - & - & - & 25.57 & 17.68 & \underline{62.38} \\
& Qwen2.5-32B$^{\triangle\dagger}$ & 32.25 & 20.24 & \textbf{88.59} & - & - & - & 28.03 & 19.36 & \textbf{68.55} \\
\midrule
\multirow{3}{*}{RepoNavigator} & Qwen2.5-7B$^{\dagger}$ & 51.63 & 53.83 & 50.62 & - & - & - & 27.49 & 30.34 & 26.69 \\
& Qwen2.5-14B$^{\dagger}$ & 58.90 & 58.97 & 61.60 & - & - & - & 29.23 & 30.08 & 31.02 \\
& Qwen2.5-32B$^{\dagger}$ & 67.75 & 70.76 & 67.29 & - & - & - & 34.09 & 37.19 & 33.71 \\
\midrule
& Qwen3-1.7B & 2.40 & 2.09 & 3.60 & 0.37 & 0.32 & 0.60 & 0.34 & 0.32 & 0.50 \\
& Qwen3-4B-Instruct & 49.73 & 49.69 & 53.34 & 19.32 & 19.86 & 20.15 & 13.27 & 14.17 & 13.74 \\
& Qwen3-14B & 43.13 & 36.49 & 71.20 & 22.86 & 20.40 & 33.04 & 16.08 & 14.51 & 23.58 \\
& Qwen3-32B (Thinking) & 62.91 & 59.87 & 73.63 & 34.69 & 33.85 & 39.46 & 23.99 & 24.22 & 26.86 \\
\rowcolor{lightgolden}\cellcolor{white}
& \textbf{\methodname-1.7B-RFT}$^{\ddagger}$ & 46.60 & 48.60 & 45.82 & 29.79 & 31.60 & 29.13 & 23.04 & 25.30 & 22.32 \\
\rowcolor{lightgolden}\cellcolor{white}
& \textbf{\methodname-1.7B}$^{\ddagger\dagger}$ & 55.46 & 58.40 & 54.27 & 36.45 & 39.37 & 35.46 & 28.22 & 31.77 & 27.18 \\

\rowcolor{lightgolden}\cellcolor{white}
& \textbf{\methodname-4B}$^{\dagger}$ & \underline{68.52} & \textbf{71.53} & 67.74 & \underline{45.97} & \underline{49.70} & \underline{44.97} & \underline{36.78} & \underline{40.71} & 35.72 \\
\rowcolor{lightgolden}\multirow{-8}{*}{OpenHands-Bash} \cellcolor{white}
& \textbf{\methodname-14B}$^{\dagger}$ & \textbf{68.57} & \underline{71.00} & 68.69 & \textbf{50.88} & \textbf{53.71} & \textbf{50.88} & \textbf{40.32} & \textbf{43.74} & 40.27 \\
\bottomrule
\end{tabular}%
}
\label{tab:swe_bench_results}
\end{table*}

%% file: Sections/5_results.tex
\input{Tables/results_pro}

\input{Tables/results_lite}

\section{Results}\label{sec:results}

This section describes the experimental results of \methodname and compares them with various competitive baselines (\S\ref{sec:baselines}). We present the summary of our results for all three benchmarks in Table \ref{tab:swe_bench_summary}. We also present detailed results for SWE-Bench Verified in Table \ref{tab:swe_bench_results}, SWE-Bench Pro in Table \ref{tab:swe_bench_pro_results}, and SWE-Bench Lite in Table \ref{tab:swe_bench_lite_results}. 

Across all three evaluation benchmarks, \methodname models (using OpenHands-Bash) achieve superior or competitive localization performance over \textbf{8-18$\times$ larger} base and post-trained LLMs using various agent scaffolds. Furthermore, \methodname often narrows the performance gap with, and sometimes even outperforms closed-source LLMs. We present key insights below and primarily compare methods using F1 score as it captures both precision and recall.

\subsection{\methodname substantially outperforms base LLMs of similar and larger sizes with OpenHands-Bash}
Across all evaluation benchmarks and localization granularities, \methodname LLMs \emph{significantly} outperform their corresponding base models when using the OpenHands-Bash scaffold. \methodname-1.7B achieves absolute improvements in F1 score over its base model of \textbf{40-54\%} at the file level, \textbf{25-41\%} at the module level, and \textbf{18-28\%} at the function level. Similarly, \methodname-4B yields absolute gains of \textbf{14-19\%}, \textbf{25-38\%}, and \textbf{21-31\%} in file, module, and function-level F1 scores, respectively. Finally, \methodname-14B achieves substantial absolute improvements of \textbf{23-34\%} in file-level F1, \textbf{25-39\%} in module-level F1, and \textbf{20-33\%} in function-level F1 scores.

Secondly, \methodname models demonstrate exceptional parameter efficiency and consistently outperform significantly larger base LLMs using the OpenHands-Bash scaffold. \methodname-1.7B outperforms the \textbf{8$\times$} larger Qwen3-14B with absolute gains in F1 score of \textbf{11-18\%} for files, \textbf{13-21\%} for modules, and \textbf{10-15\%} for functions. When compared against the \textbf{18$\times$} larger Qwen3-32B (Thinking) model, \methodname-1.7B remains competitive, surpassing its F1 score by \textbf{2-4\%} and \textbf{3-6\%} for modules and functions respectively, while trailing by \textbf{2-7\%} for files. \methodname-4B consistently outperforms \textbf{8$\times$} larger Qwen3-32B for all benchmarks, with absolute improvements in F1 score of \textbf{5-8\%}, \textbf{11-16\%}, and \textbf{13-17\%} for files, modules, and functions respectively. Finally, \methodname-14B further widens the gap over Qwen3-32B, surpassing its F1 scores by \textbf{6-13\%} for files, \textbf{15-22\%} for modules, and \textbf{16-21\%} for functions.

\subsection{\methodname outperforms larger base and post-trained LLMs using complex scaffolds}

\methodname demonstrates superior or competitive performance over larger base and post-trained LLMs that utilize complex specialized scaffolds.

\textbf{Comparisons with larger base LLMs using complex scaffolds:} \methodname models consistently outperform Qwen2.5-32B-Instruct with the OrcaLoca scaffold~\citep{yu2025orcalocallmagentframework} on SWE-Bench Verified. \methodname-4B and \methodname-14B \emph{significantly} exceed its F1 scores by \textbf{10\%} for files and \textbf{8-11\%} for functions. Impressively, our \textbf{18$\times$ smaller} \methodname-1.7B remains  competitive, trailing Qwen2.5-32B with OrcaLoca by \textbf{$<$ 0.5\%} in function-level F1 and by \textbf{2.65\%} in file-level F1. On SWE-bench Pro, \methodname models exhibit more pronounced gains over Qwen2.5-32B-Instruct with the CoSIL scaffold~\citep{liu2025cosil}. \methodname-1.7B, 4B, and 14B achieve absolute F1 gains of \textbf{20-33\%} for files and \textbf{11-21\%} for functions.

\textbf{Comparisons with larger post-trained LLMs using complex scaffolds:} We primarily compare \methodname with RepoNavigator~\citep{zhang2025one} as it is our strongest post-training baseline. It is also closer to our work than  RepoSearcher~\citep{ma2025toolintegratedreinforcementlearningrepo} as it is trained without relying on closed-source LLMs and can predict a variable number of locations (\S\ref{sec:baselines}).

For both SWE-bench Verified and Pro, \methodname models frequently achieve superior or competitive performance over \textbf{3-8$\times$ larger} RepoNavigator models. For both benchmarks, \methodname-1.7B achieves slightly better performance than RepoNavigator-7B  with a \textbf{1-4\%} higher F1 score for both files and functions. Furthermore, \methodname-4B consistently exceeds the performance of RepoNavigator-14B across both benchmarks, with absolute F1 improvements of \textbf{2-10\%} for files and \textbf{8-11\%} for functions. Comparison with RepoNavigator-32B reveals distinct trends across benchmarks. On SWE-bench Verified, \methodname-4B and \methodname-14B achieve better function-level F1 by \textbf{3-6\%} with a very similar file-level F1. On SWE-bench Pro, \methodname-4B and \methodname-14B outperform RepoNavigator-32B by \textbf{8\%} in function-level F1, whereas the 32B baseline has \textbf{4-6\%} better file-level F1 scores.

\subsection{\methodname narrows the performance gap with closed-source LLMs}
On SWE-bench Verified, \methodname-4B and \methodname-14B achieve superior results compared to GPT-5-Chat with RepoNavigator, outperforming it by \textbf{9\%} in file-level F1 and \textbf{5-9\%} in function-level F1. Moreover, \methodname-4B and \methodname-14B achieve impressive function-level localization performance surpassing the function-level F1 score of Claude-3.7-Sonnet with RepoNavigator by \textbf{5-8\%}. These results are significant given that \methodname uses a bash-only agent as opposed to the specialized RepoNavigator scaffold.

Interestingly, Claude-Sonnet-4.5 achieves stronger performance with our bash-only agent over RepoNavigator, with significant improvement in function-level F1 score by \textbf{18\%} and a slightly better file-level F1 score (2\%). Similarly GPT-5 with our agent outperforms Claude 3.7 Sonnet with the RepoNavigator scaffold, with absolute improvements in F1 of \textbf{23\%} for functions and \textbf{5\%} for files. These results suggest that designing specialized scaffolds may not always lead to better performance with frontier models and can even degrade model performance. Prior agent frameworks like OpenHands-Versa \citep{soni2025codingagentsmultimodalbrowsing} and mini SWE-Agent~\citep{mini_swe_agent_repo,yang2024sweagentagentcomputerinterfacesenable} have reported similar findings for other task domains.

For all benchmarks, GPT-5 and Claude-Sonnet-4.5 consistently outperform \methodname when using the OpenHands-Bash scaffold with an additional reminder (\S\ref{sec:baselines}). A surprising finding is the sensitivity of these frontier LLMs to prompt engineering; without the additional reminder for prediction submission, the performance of both models drops to almost \textbf{zero} on all benchmarks. Interestingly, our preliminary experiments do not reveal similar problems with GPT-4o despite being an older proprietary LLM.

%% file: Tables/results_pro.tex
\begin{table*}[!ht]
\caption{Results on \textbf{SWE-Bench Pro}. \methodname outperforms base and post-trained LLMs up to \textbf{8$\times$ larger} across multiple scaffolds, narrowing the gap with closed-source LLMs. Qwen2.5 results use instruct variants. $^{rem}$ reminds the LLM to submit answer (\S\ref{sec:baselines}), $^\dagger$trained with RL (GRPO/GSPO); $^\ddagger$RFT from \methodname-14B. The highest metric is \textbf{bold-faced} and 2$^{nd}$ highest metric is \underline{underlined}.}
\centering
\renewcommand{\arraystretch}{1.2} 
\setlength{\tabcolsep}{4.0pt} 
\resizebox{\textwidth}{!}{%
\begin{tabular}{lc|ccccccccc}
\toprule
\multirow{2}{*}{\textbf{Scaffold}} & \multirow{2}{*}{\textbf{LLM}} & \multicolumn{3}{c}{\textbf{File-level}} & \multicolumn{3}{c}{\textbf{Module-level}} & \multicolumn{3}{c}{\textbf{Function-level}} \\
& & F1 & Prec. & Rec. & F1 & Prec. & Rec. & F1 & Prec. & Rec. \\
\midrule
\rowcolor{headergray}
\multicolumn{11}{c}{\textbf{Closed-Source LLMs}} \\
\midrule
\multirow{2}{*}{OpenHands-Bash} & GPT-5 & 0.00 & 0.00 & 0.00 & 0.00 & 0.00 & 0.00 & 0.00 & 0.00 & 0.00 \\
& Claude-Sonnet-4.5 & 0.00 & 0.00 & 0.00 & 0.00 & 0.00 & 0.00 & 0.00 & 0.00 & 0.00 \\
\multirow{2}{*}{OpenHands-Bash$^{rem}$} & GPT-5 & \underline{61.18} & \underline{69.10} & \underline{62.06} & \underline{42.20} & \underline{52.86} & \underline{39.63} & \underline{35.86} & \underline{48.91} & \underline{32.65} \\
& Claude-Sonnet-4.5 & \textbf{64.75} & \textbf{75.39} & \textbf{64.49} & \textbf{48.54} & \textbf{61.75} & \textbf{45.39} & \textbf{42.26} & \textbf{58.14} & \textbf{38.06} \\
\midrule
\rowcolor{headerblue}
\multicolumn{11}{c}{\textbf{Open-Source LLMs}} \\
\midrule
RepoSearcher & \multirow{4}{*}{Qwen2.5-32B}& 
3.81 & 2.52 & 9.00 & 
- & - & - & 
2.31 & 2.46 & 2.52 \\
LocAgent & & 
19.77 & 0.38 & 25.73 & 
- & - & - & 
4.30 & 0.17 & 8.72 \\
Agentless &  & 
20.07 & 13.89 & 43.07 & 
- & - & - & 
7.98 & 7.31 & 11.08 \\
CoSIL & & 
20.95 & 14.03 & 48.87 & 
- & - & - & 
7.67 & 6.00 & 14.03 \\
\midrule
\multirow{3}{*}{RepoNavigator} & Qwen2.5-7B$^{\dagger}$ & 
39.74 & 48.13 & 36.36 & 
- & - & - & 
14.29 & 21.26 & 12.33 \\
& Qwen2.5-14B$^{\dagger}$ & 
49.72 & 58.64 & 46.85 & 
- & - & - & 
18.06 & 25.25 & 16.05 \\
& Qwen2.5-32B$^{\dagger}$ & 
\textbf{57.57} & 68.69 & \underline{53.49} & 
- & - & - & 
20.72 & 29.44 & 18.13 \\
\midrule
& 
Qwen3-1.7B & 
0.73 & 0.72 & 1.13 & 
0.00 & 0.00 & 0.00 & 
0.00 & 0.00 & 0.00 \\
& Qwen3-4B-Instruct & 
36.96 & 44.42 & 35.59 & 
11.78 & 17.46 & 10.19 & 
8.12 & 12.16 & 7.01 \\
& Qwen3-14B & 
30.08 & 28.48 & 48.22 & 
11.87 & 13.82 & 14.21 & 
8.20 & 9.97 & 9.92 \\
& Qwen3-32B (Thinking) & 
46.85 & 49.65 & \textbf{54.55} & 
21.94 & 27.18 & 22.62 & 
12.31 & 17.82 & 11.93 \\
\rowcolor{lightgolden}\multirow{-2}{*}{OpenHands-Bash} \cellcolor{white}
& \textbf{\methodname-1.7B-RFT}$^{\ddagger}$ & 34.54 & 47.74 & 30.22 & 22.43 & 36.53 & 18.59 & 16.07 & 29.04 & 13.02 \\
\rowcolor{lightgolden}\cellcolor{white}
& \textbf{\methodname-1.7B}$^{\ddagger\dagger}$ & 40.96 & 56.52 & 35.91 & 25.27 & 40.66 & 20.83 & 18.24 & 32.08 & 14.72 \\
\rowcolor{lightgolden}\cellcolor{white} & 
\textbf{\methodname-4B}$^{\dagger}$ & 
51.77 & \textbf{68.98} & 46.16 & 
\underline{36.97} & \textbf{56.05} & \underline{31.13} & 
\textbf{29.03} & \textbf{48.65} & \underline{23.73} \\
\rowcolor{lightgolden}\cellcolor{white}
& \textbf{\methodname-14B}$^{\dagger}$ & 
\underline{53.63} & \underline{68.81} & 48.81 & 
\textbf{37.13} & \underline{53.80} & \textbf{32.02} & 
\underline{28.74} & \underline{46.09} & \textbf{23.76} \\
\bottomrule
\end{tabular}%
}
\label{tab:swe_bench_pro_results}
\end{table*}

%% file: Tables/results_lite.tex
\begin{table*}[!ht]
\caption{Results on \textbf{SWE-Bench Lite}. \methodname achieves better/comparable results over base LLMs upto \textbf{18$\times$ larger}, narrowing the gap with closed-source LLMs. Qwen2.5 results use instruct models. $^{rem}$ adds a submission reminder (\S\ref{sec:baselines}), $^\dagger$trained with GRPO/GSPO; $^\ddagger$RFT from \methodname-14B; $^\triangle$RFT from Claude-3.7-Sonnet. The best metric is \textbf{bold-faced} and 2$^{nd}$ best metric is \underline{underlined}.}
\centering
\renewcommand{\arraystretch}{1.2}
\setlength{\tabcolsep}{4.0pt}
\resizebox{\textwidth}{!}{%
\begin{tabular}{lc|ccccccccc}
\toprule
\multirow{2}{*}{\textbf{Scaffold}} & \multirow{2}{*}{\textbf{LLM}} & \multicolumn{3}{c}{\textbf{File-level}} & \multicolumn{3}{c}{\textbf{Module-level}} & \multicolumn{3}{c}{\textbf{Function-level}} \\
& & F1 & Prec. & Rec. & F1 & Prec. & Rec. & F1 & Prec. & Rec. \\
\midrule
\rowcolor{headergray}
\multicolumn{11}{c}{\textbf{Closed-Source LLMs}} \\
\midrule
LocAgent & Claude-3.5-Sonnet & 31.39 & 18.83 & \textbf{94.16} & 29.91 & 18.18 & \textbf{86.98} & 27.53 & 17.08 & \textbf{76.61} \\
\midrule
\multirow{2}{*}{OpenHands-Bash} & GPT-5 & 1.09 & 1.09 & 1.09 & 1.09 & 1.09 & 1.09 & 1.09 & 1.09 & 1.09 \\
& Claude-Sonnet-4.5 & 0.36 & 0.36 & 0.36 & 0.36 & 0.36 & 0.36 & 0.36 & 0.36 & 0.36 \\
\multirow{2}{*}{OpenHands-Bash$^{rem}$} & GPT-5 & \underline{77.77} & \underline{75.73} & 82.48 & \underline{67.86} & \underline{66.09} & 72.63 & \textbf{61.12} & \underline{59.43} & \underline{67.21} \\
& Claude-Sonnet-4.5 & \textbf{81.87} & \textbf{80.17} & \underline{85.40} & \textbf{69.62} & \textbf{68.87} & \underline{72.87} & \underline{61.11} & \textbf{61.72} & 63.72 \\
\midrule
\rowcolor{headerblue}
\multicolumn{11}{c}{\textbf{Open-Source LLMs}} \\
\midrule
& 
Qwen3-1.7B & 2.16 & 1.96 & 2.92 & 0.36 & 0.36 & 0.36 & 0.00 & 0.00 & 0.00 \\
& Qwen3-4B-Instruct & 47.41 & 43.70 & 55.47 & 14.50 & 13.90 & 16.24 & 8.72 & 8.67 & 9.61 \\
& Qwen3-14B & 38.63 & 31.30 & \underline{71.90} & 20.10 & 16.74 & 31.57 & 11.73 & 9.88 & 18.13 \\
& Qwen3-32B (Thinking) & 58.98 & 54.26 & 71.53 & 37.20 & 34.64 & 43.98 & 23.76 & 23.11 & 26.89 \\
\rowcolor{lightgolden}
\multirow{-2}{*}{OpenHands-Bash} \cellcolor{white}
& \textbf{\methodname-1.7B-RFT}$^{\ddagger}$ & 45.99 & 45.99 & 45.99 & 34.79 & 35.22 & 34.67 & 24.51 & 25.18 & 24.39 \\
\rowcolor{lightgolden}\cellcolor{white}
& \textbf{\methodname-1.7B}$^{\ddagger\dagger}$ & 56.57 & 56.57 & 56.57 & 41.24 & 41.61 & 41.24 & 27.07 & 28.28 & 26.82 \\
\rowcolor{lightgolden}
\cellcolor{white} & \textbf{\methodname-4B}$^{\dagger}$ & \underline{67.03} & \underline{66.61} & 67.88 & \underline{53.10} & \underline{53.47} & \underline{53.65} & \underline{39.87} & \underline{41.59} & \underline{39.96} \\
\rowcolor{lightgolden}
\cellcolor{white} & \textbf{\methodname-14B}$^{\dagger}$ & \textbf{71.84} & \textbf{71.17} & \textbf{73.36} & \textbf{59.23} & \textbf{59.18} & \textbf{60.28} & \textbf{44.43} & \textbf{45.59} & \textbf{45.13} \\
\bottomrule
\end{tabular}%
}
\label{tab:swe_bench_lite_results}
\end{table*}

%% file: Sections/6_analysis.tex
\section{Analysis}
This section presents detailed analysis offering more insights into the \methodname recipe. We discuss the advantages of effective code localization on issue resolution (\S\ref{sec:downstream_swebench}) and analyze evolving tool-use behaviors during RL (\S\ref{sec:tool_use_behaviour}).

\subsection{Does Effective Code Localization Improve Issue Resolution?}\label{sec:downstream_swebench}
\input{Figures/heatmap_14B}

\input{Tables/downstream_swebench_performance}

We demonstrate that augmenting issue resolution agents with relevant code locations retrieved by \methodname improves downstream performance on SWE-Bench Verified when using the OpenHands Agent SDK~\citep{wang2025openhandssoftwareagentsdk}. We consider three settings: (1) a vanilla baseline without localization, (2) augmenting the agent with locations retrieved by \methodname-14B, and (3) augmenting the agent with oracle locations parsed from the gold patch (\S\ref{sec:data_creation}). In settings (2) and (3), the user prompt is modified to include the names of localized files, modules, and functions (Appendix~\ref{appendix:prompts_swe_bench}). Experiments are conducted with Qwen3-4B-Instruct-2507 and Qwen3-Coder-30B-A3B-Instruct using a 128K context limit, allowing upto 100 steps, and decoding hyperparameters recommended by Qwen3 developers. We compare the three methods in terms of performance using issue resolution rate (\% of issues successfully fixed), and efficiency using instance-level averages of the number of steps per trajectory and the total input and output tokens accumulated across all turns. Table~\ref{tab:downstream_swe_bench} presents the detailed results.

For Qwen3-4B-Instruct, improved localization quality leads to substantial gains in both performance and efficiency: augmenting the agent with locations retrieved by \methodname-14B increases the resolution rate by \textbf{3.8\%}, while reducing trajectory length by \textbf{2.18} steps, and input and output tokens by \textbf{17.46\%} and \textbf{6.71\%} respectively. Similar efficiency gains are observed for Qwen3-Coder-30B-A3B-Instruct, with \textbf{2.89} fewer steps and \textbf{5.33\%} fewer input tokens and \textbf{2.00\%} fewer output tokens, while achieving comparable performance with a small improvement (+0.80\%). For both models, replacing \methodname predictions with oracle locations yields further performance gains (\textbf{+6.00\%} for 30B and \textbf{+2.40\%} for 4B). The 30B model additionally benefits with improved efficiency, with \textbf{1.37} fewer steps and reductions of \textbf{10.12\%} and \textbf{5.60\%} in input and output tokens. Overall, these results indicate that continued improvements in code localization will likely further enhance both the performance and efficiency of issue resolution agents.

\subsection{How does the tool-use behaviour of \methodname evolve during RL?}\label{sec:tool_use_behaviour}

To analyze the evolving tool-use behavior of our 4B and 14B models during training, we plot the distribution of Unix command-line utilities invoked by the agent. We parse the \texttt{Terminal} tool calls from the rollouts logged during training to extract the Unix commands. We consider rollouts sampled from model checkpoints at every 10 training steps. Command usage is aggregated across these checkpoints to identify the top-8 most frequently used utilities. For each checkpoint, we then compute the relative frequency with which these utilities are invoked across all rollouts logged for that training step. Figure~\ref{fig:merged_tool_use} presents the relative usage distribution of these utilities as training proceeds.

We observe a clear convergence in tool usage as training proceeds. While the 14B model initially invokes various utilities (e.g., \texttt{grep}, \texttt{find}, \texttt{wc}, and \texttt{cat}), after approximately 200 training steps it mostly relies \emph{only} on two commands: \texttt{ripgrep} (\texttt{rg}) and \texttt{sed}. Similarly, while the 4B model initially uses a many utilities, it eventually converges to primarily using \texttt{ripgrep}, \texttt{sed}, \texttt{cat}, \texttt{find} and \texttt{xargs}. Appendix~\ref{appendix:traj_examples} presents example trajectories illustrating how the 4B and 14B LLMs use these utilities for code localization. These findings suggest that effective localization can be achieved using an even simpler scaffold with only a small subset of Unix utilities without requiring access to the entire Unix command-line interface which is crucial in security-sensitive deployments.

%% file: Figures/heatmap_14B.tex
\begin{figure*}[!h]
    \centering
    \newcommand{\figwidth}{0.49\linewidth}
    \newcommand{\figheight}{4.5cm} %
        \begin{subfigure}[b]{\figwidth}
        \centering
        \includegraphics[width=\linewidth, height=\figheight]{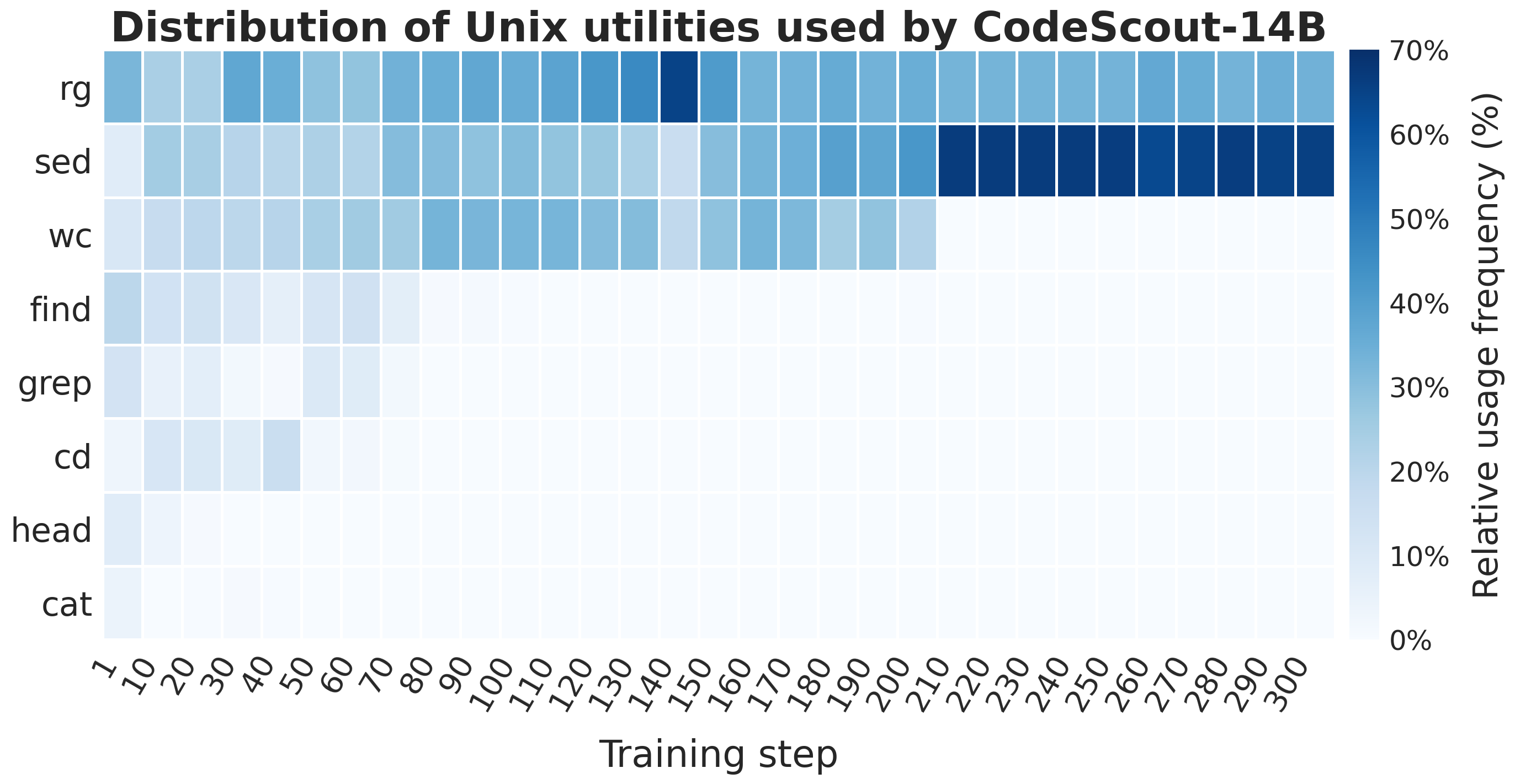}
        \caption{Tool use for \methodname-14B}
        \label{fig:tool_use_14b_sub}
    \end{subfigure}
    \hfill
    \begin{subfigure}[b]{\figwidth}
        \centering
        \includegraphics[width=\linewidth, height=\figheight]{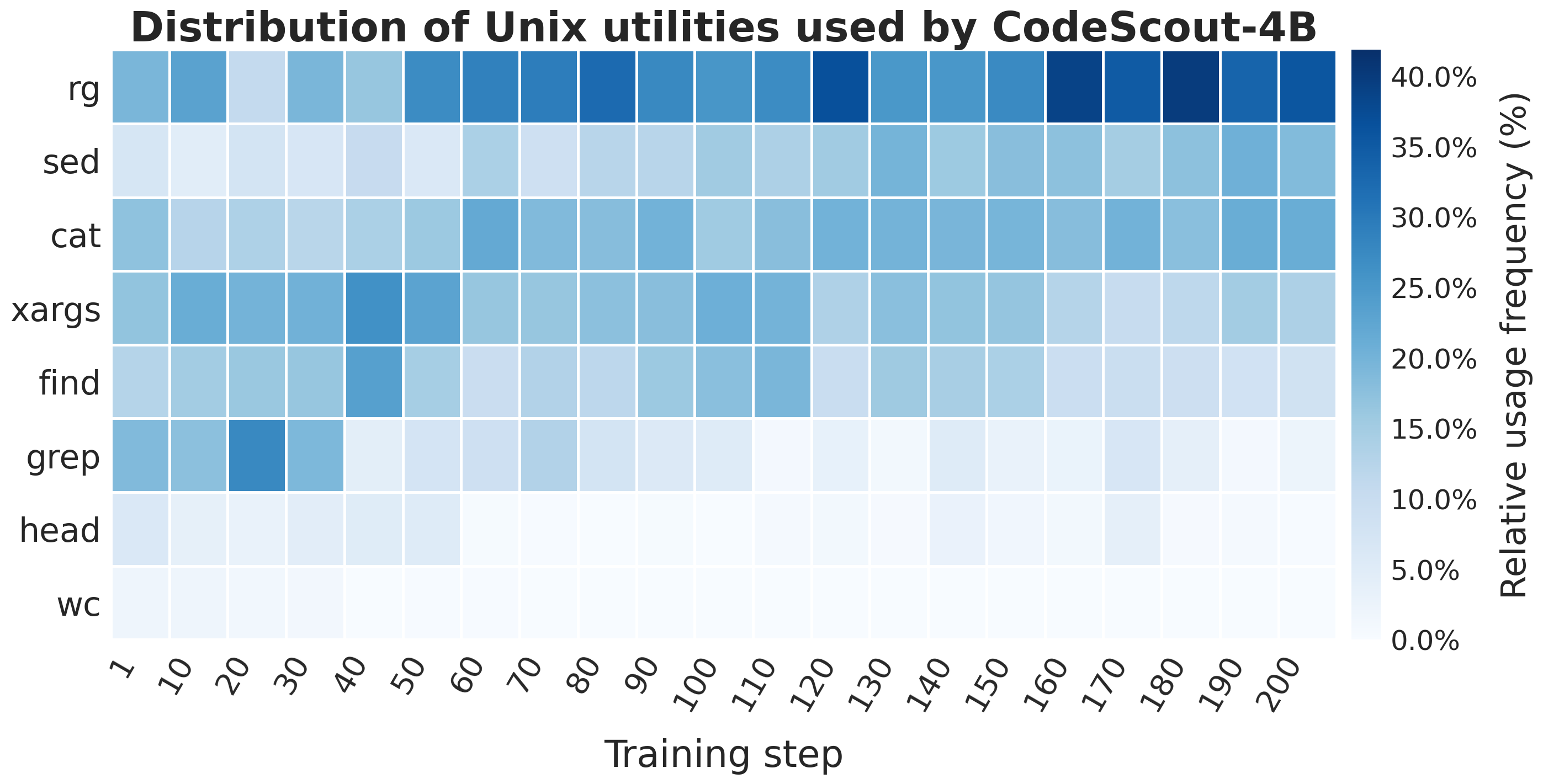}
        \caption{Tool use for \methodname-4B}
        \label{fig:tool_use_4b_sub}
    \end{subfigure}

    \caption{Distribution of the top-8 most frequent Unix commands used by \methodname at different training stages. While both LLMs initially use a broad range of Unix utilities, they eventually use a very limited set of commands as training proceeds. \methodname-14B \emph{only} uses ripgrep (rg) and sed, whereas \methodname-4B mainly uses rg, cat, sed, and xargs.}
    \label{fig:merged_tool_use}
\end{figure*}

%% file: Tables/downstream_swebench_performance.tex
\begin{table}[!t]
\caption{Issue resolution performance on SWE-Bench Verified: augmenting the agent with relevant code locations achieves higher resolution rate while reducing number step count and token usage. The best metric is \textbf{bold-faced} and 2$^{nd}$ best metric is \underline{underlined}.}
\centering
\renewcommand{\arraystretch}{1.2}
\setlength{\tabcolsep}{2.3pt}
\resizebox{\columnwidth}{!}{%
\begin{tabular}{lcccc} %
\toprule
\makecell[l]{\textbf{Localization}\\\textbf{Approach}} & \makecell{\textbf{Resolution}\\\textbf{Rate $\uparrow$}} & \makecell{\textbf{Avg. \# }\\\textbf{Steps $\downarrow$}} & \makecell{\textbf{Avg. Input}\\\textbf{Tokens $\downarrow$}} & \makecell{\textbf{Avg. Output}\\\textbf{Tokens $\downarrow$}} \\
\midrule
\rowcolor{blue!15!white}\multicolumn{5}{c}{\textbf{Qwen3-4B-Instruct}} \\
\midrule
None (Vanilla) & 13.40\% & \underline{16.09} & 344.37K & \underline{2.98K} \\
\methodname-14B & \underline{17.20\%} & \textbf{13.91} & \textbf{284.26K} & \textbf{2.78K} \\
Oracle & \textbf{19.60\%} & 16.41 & \underline{327.75K} & 3.17K \\
\midrule
\rowcolor{pink!40!white}\multicolumn{5}{c}{\textbf{Qwen3-Coder-30B-A3B-Instruct}} \\ %
\midrule
None (Vanilla) & 45.20\% & 51.00 & 1596.71K & 13.49K \\
\methodname-14B & \underline{46.00\%} & \underline{48.11} & \underline{1511.53K} & \underline{13.22K} \\
Oracle & \textbf{52.00\%} & \textbf{46.74} & \textbf{1358.60K} & \textbf{12.48K} \\
\bottomrule
\end{tabular}%
}
\label{tab:downstream_swe_bench}
\end{table}

%% file: Sections/8_conclusion.tex
\section{Conclusion}
We present \methodname: a fully open-source reinforcement learning recipe for training code search agents, and release the \methodname model family that demonstrates state-of-the-art localization performance without relying on complex scaffolds dependent on programming language-specific static analysis tools as done by prior work. By leveraging a simple agent scaffold equipped with \emph{just} a terminal tool, \methodname achieves remarkable localization efficacy. Our experiments on SWE-Bench Verified, Pro, and Lite show that models trained with our recipe not only outperform significantly larger open-source models (both pre-trained and post-trained variants) across various complex scaffolds, but also close the performance gap with some proprietary closed-source LLMs. Our analysis reveals the advantages of augmenting issue resolution agents with relevant localized context both in terms of efficiency and performance. We also study the fine-grained changes in model behaviour during RL in terms of the command-line utilities leveraged for localization, revealing the feasibility of developing even simpler scaffolds with strong performance on this task. Appendix~\ref{appendix:future_work} includes additional discussion on limitations and directions for future work. We publicly release our code, data, and model weights for future work to build upon, and leverage our infrastructure to train effective LLM-based code agents.

%% file: Sections/9_impact_statement.tex
\section*{Impact Statement}
Our work studies how to train language models to perform repository-level code localization, a key capability underlying many coding agent tasks like automated issue resolution. Advances in code localization will help strengthen automated software engineering systems that rely on accurate code retrieval. Progress in coding agents has significant societal implications as they can reshape the software engineering profession, reducing the demand for certain types of routine programming tasks contributing towards workforce displacement. We advocate for the responsible development and deployment of coding agents, with appropriate safeguards and human oversight to mitigate real-world harm.

\section*{Acknowledgements}
We would like to thank Xingyao Wang, Valerie Chen, Jim White and Paul Cuciureanu for the insightful discussions and feedback. We also thank Modal and CMU Flame Center for providing access to GPU resources for our training runs. This work is generously supported by grants from IBM, Apple, and Amazon. The views, opinions, and findings expressed in this work are solely those of the authors and do not necessarily reflect those of the funding agencies.

%% file: Sections/appendix.tex
\appendix
\onecolumn

\section{Prompts and Tool Definitions for OpenHands-Bash}\label{appendix:prompts}
Figures \ref{fig:system_prompt1},~\ref{fig:system_prompt2} describe the system prompt, and Figure \ref{fig:user_prompt} includes the user prompt, used for all LLMs (including \methodname) trained or evaluated with OpenHands-Bash (\S\ref{sec:agent_scaffold}). The system prompt specifies the \texttt{max\_turns} to be 6 for \methodname-4B and its base model, and to 4 for all other LLMs. Table~\ref{tab:pythonic_tool_schemas} provides the schemas for the tools used by OpenHands-Bash.
\begin{figure*}[!h]
    \centering
\begin{tcolorbox}[colback=blue!6!white, colframe=blue!75!black, title=System Prompt for the OpenHands-Bash agent used by \methodname (continued on next page)]
\begin{lstlisting}
You are a specialized code localization agent. Your sole objective is to identify and return the files in the codebase that are relevant to the user's query.
You are given access to the codebase in a linux file system.

## PRIMARY DIRECTIVE
- Find relevant files, do NOT answer the user's query directly
- Prioritize precision: every file you return should be relevant
- You have up to {{ max_turns }} turns to explore and return your answer

## TOOL USAGE REQUIREMENTS

### bash tool (REQUIRED for search)
- You MUST use the bash tool to search and explore the codebase
- Execute bash commands like: rg, grep, find, ls, cat, head, tail, sed
- Use parallel tool calls: invoke bash tool up to 5 times concurrently in a single turn
- NEVER exceed 5 parallel tool calls per turn
- Common patterns:
  * `rg "pattern" -t py` - search for code patterns
  * `rg --files | grep "keyword"` - find files by name
  * `cat path/to/file.py` - read file contents
  * `find . -name "*.py" -type f` - locate files by extension
  * `wc -l path/to/file.py` - count lines in a file
  * `sed -n '1,100p' path/to/file.py` - read lines 1-100 of a file
  * `head -n 100 path/to/file.py` - read first 100 lines
  * `tail -n 100 path/to/file.py` - read last 100 lines

### Reading Files (CRITICAL for context management)
- NEVER read entire large files with `cat` - this will blow up your context window
- ALWAYS check file size first: `wc -l path/to/file.py`
- For files > 100 lines, read in chunks:
  * Use `sed -n '1,100p' file.py` to read lines 1-100
  * Use `sed -n '101,200p' file.py` to read lines 101-200
  * Continue with subsequent ranges as needed (201-300, 301-400, etc.)
- Strategic reading approach:
  * Read the first 50-100 lines to see imports and initial structure
  * Use `rg` to find specific patterns and their line numbers
  * Read targeted line ranges around matches using `sed -n 'START,ENDp'`
  * Only read additional chunks if the initial sections are relevant

### Submitting Your Answer (REQUIRED)

When you have identified all relevant locations, you MUST use the `localization_finish` tool to submit your results.

**When to include what:**
1. If the required modifications belong to a specific function that belongs to a class, provide the file path, class name, and function name.
2. If the required modification belongs to a function that is not part of any class, provide the file path and function name.
\end{lstlisting}
\end{tcolorbox}
\caption{System prompt for the OpenHands-Bash agent scaffold used by \methodname (continued on next page).}
\label{fig:system_prompt1}
\end{figure*}

\begin{figure*}[!ht]
    \centering
\begin{tcolorbox}[colback=blue!6!white, colframe=blue!75!black, title=System Prompt for the OpenHands-Bash agent used by \methodname]
\begin{lstlisting}
3. If the required modification does not belong to any specific class or a function (e.g. global variables, imports, new class, new global function etc.), it is sufficient to provide only the file path.
4. If the required modification belongs to a class (e.g. adding a new method to a class), provide the file path and class name.

## SEARCH STRATEGY

1. **Initial Exploration**: Cast a wide net
   - Search for keywords, function names, class names
   - Check file names and directory structure
   - Use up to 3 parallel bash calls to explore multiple angles
   - Check file sizes with `wc -l` before reading
   - Read promising files in chunks (lines 1-100) to verify relevance

2. **Deep Dive**: Follow the most promising leads
   - Use up to 3 parallel bash calls to investigate further
   - Read files in chunks to confirm they address the query
   - Use `rg` with line numbers to locate specific code, then read those ranges
   - Start eliminating false positives

3. **Final Verification**: Confirm your location list and terminate execution by calling the `localization_finish` tool

## CRITICAL RULES
- NEVER exceed 5 parallel bash tool calls in a single turn
- ALWAYS use the `localization_finish` tool after identifying all relevant locations
- ALWAYS use bash tool to search (do not guess file locations)
- NEVER read entire large files - always read in chunks (100-line ranges)
- Check file size with `wc -l` before reading
- Read file contents in chunks to verify relevance before including them
- Return file paths as they appear in the repository. Do not begin the path with "./"
- Aim for high precision (all files relevant) and high recall (no relevant files missed)
- Class and function names are OPTIONAL - only include when changes are at that level

## EXAMPLE OUTPUT BEHAVIOUR
Here are some examples of how to format your output when calling the `localization_finish` tool:
- src/parsers/parser.py requires changes to imports, a function parse_data which belongs to the class DataParser, and another function __str__ inside the same class. This should be represented as three separate entries: one with just the file path and one each for the two functions parse_data and __str__ with file path, class name, and function name.
- src/user.py requires changes to a global function get_user outside of any class. This should be represented as a single entry with file path and function name.
- utils/visualizer.py requires adding new function visualize inside the class Visualizer. This should be represented as a single entry with file path and class name.
- utils/configs/default_config.py requires adding a new global function and a new class. This should be represented as a single entry with just the file path. Do NOT include class or function names for this file since multiple implementations might be possible with different function names.
\end{lstlisting}
\end{tcolorbox}
\caption{System prompt for the OpenHands-Bash scaffold used by \methodname.}
\label{fig:system_prompt2}
\end{figure*}

\begin{figure*}[!ht]
    \centering
\begin{tcolorbox}[colback=orange!6!white, colframe=orange!75!black, title=User prompt for the OpenHands-Bash agent used by \methodname]
\begin{lstlisting}
I have access to a python code repository in the directory {{ working_dir }} . Consider the following issue description:

<issue_description>
{{ instance.problem_statement }}
</issue_description>

Act as a code search agent and localize the specific files, classes or functions of code that need modification to resolve the issue in <issue_description>.

NOTE: You do not need to solve the issue, all you need to do is localize relevant code from the repository. Your output will be used to guide another agent to solve the issue.

IMPORTANT: Your output MUST follow the below rules:
1. The final output must be a tool call to the "localization_finish" tool containing relevant code locations.
2. The locations of the file path must be RELATIVE to the {{ working_dir }} directory WITHOUT any leading "./" in the output.
3. Only include those locations in your output that need modification to resolve the issue in <issue_description>. Do NOT include any locations that do not need modification.
\end{lstlisting}
\end{tcolorbox}
\caption{User prompt for the OpenHands-Bash scaffold used by \methodname.}
\label{fig:user_prompt}
\end{figure*}

\begin{table}[ht]
    \centering
    \small
    \caption{Pythonic Tool Schema for the tools used by OpenHands-Bash}
    \label{tab:pythonic_tool_schemas}
    \begin{tabular}{lllc}
        \toprule
        \textbf{Tool Name} & \textbf{Parameters} & \textbf{Python Type / Constraints} & \textbf{Required} \\
        \midrule
        \multirow{5}{*}{\texttt{terminal}} 
            & \texttt{command} & \texttt{str} & \cmark \\
            & \texttt{security\_risk} & \texttt{Literal["UNKNOWN", "LOW", "MEDIUM", "HIGH"]} & \cmark \\
            & \texttt{is\_input} & \texttt{bool} & \xmark \\
            & \texttt{timeout} & \texttt{float} & \xmark \\
            & \texttt{reset} & \texttt{bool} & \xmark \\
        \midrule
        \multirow{4}{*}{\texttt{localization\_finish}} 
            & \texttt{locations} & \texttt{List[Dict[str, Optional[str]]]} & \cmark \\
            & \quad $\hookrightarrow$ \texttt{file} & \texttt{str} & \cmark \\
            & \quad $\hookrightarrow$ \texttt{class\_name} & \texttt{str} $|$ \texttt{None} & \xmark \\
            & \quad $\hookrightarrow$ \texttt{function\_name} & \texttt{str} $|$ \texttt{None} & \xmark \\
        \bottomrule
    \end{tabular}
\end{table}

\section{Reward and Loss Curves for \methodname}\label{appendix:rl_loss_curves}
\input{Figures/codescout_14b_reward_fig}
\input{Figures/codescout_4b_reward_fig}
\input{Figures/codescout_1.7b_reward_fig}
\begin{wrapfigure}{r}{0.45\textwidth} %
    \centering
    \includegraphics[width=0.43\textwidth]{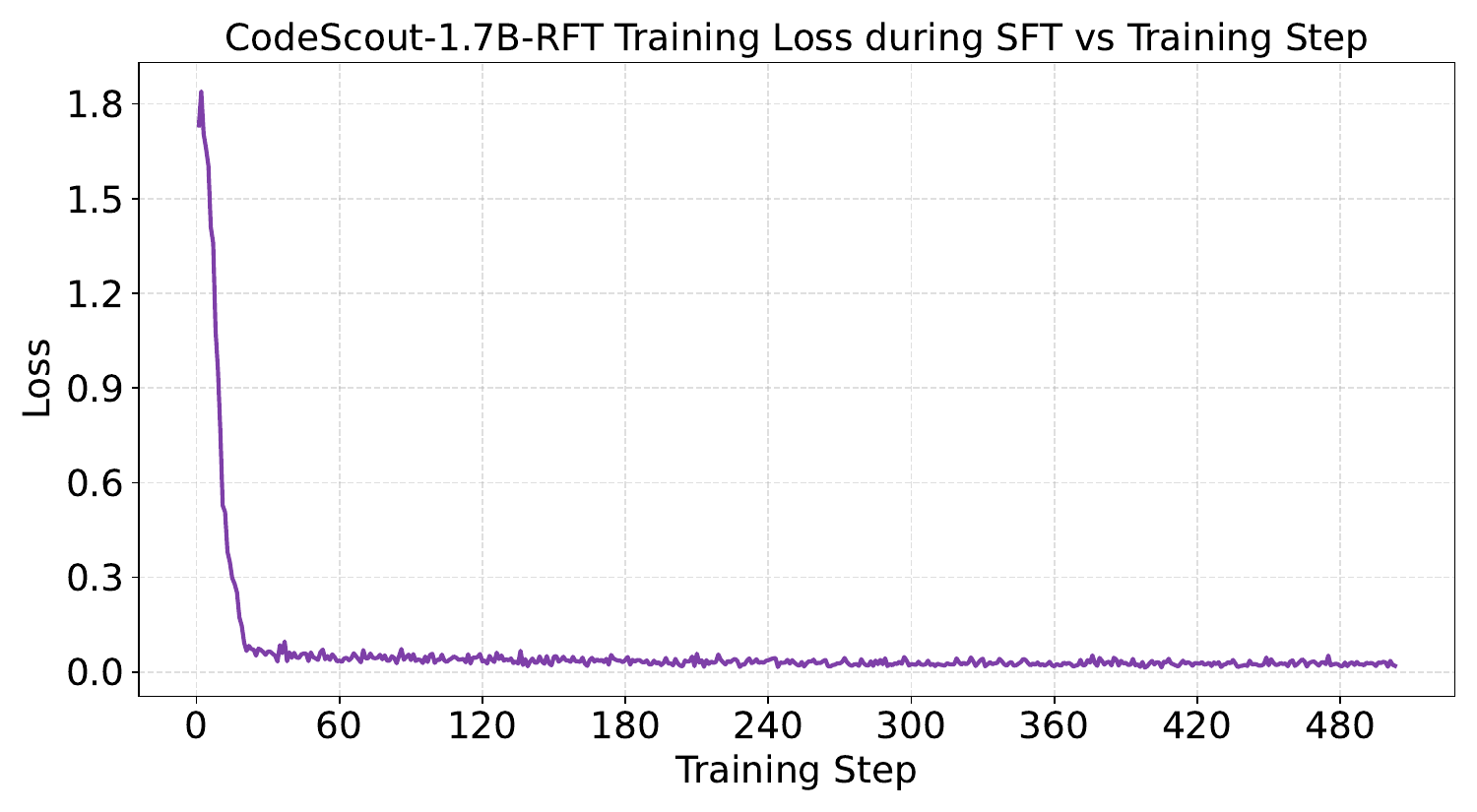} 
    \caption{Training Loss curve during rejection sampling finetuning used to train \methodname-1.7B-RFT.}
    \label{fig:codescout_1.7B_RFT}
\end{wrapfigure}
This section presents the RL reward curves and SFT loss curves from the training runs of \methodname. We present the aggregate reward (computed as the sum of the F1 scores across the three localization granularities for 4B and 1.7B, and for 14B as the sum of these three F1 scores and the auxiliary binary reward) and the individual file-level, module-level, and function-level F1 scores during training. Note that all reward curves are smoothed using a running average with a window of 16 training steps. Figures~\ref{fig:codescout_14b_reward}, \ref{fig:codescout_4b_reward}, and \ref{fig:codescout_1.7b_reward} present these reward curves for \methodname-14B, \methodname-4B, and \methodname-1.7B respectively. Furthermore, Figure~\ref{fig:codescout_1.7B_RFT} presents the loss curve for rejection sampling fine-tuning of \texttt{Qwen3-1.7B} resulting in \methodname-1.7B-RFT model.

As expected, all reward curves across all models show an approximately increasing nature (or remain constant during the later stages of training). Notably, all the reward curves for \methodname-4B and the \methodname-14B models show an increasing trend with no visible signs of saturation, indicating that training further on more GitHub issues will likely result in even stronger localization performance. However, we observe that the reward curves for \methodname-1.7B have mostly plateaued and the model no longer shows signs of further improvement with more training. Interstingly, even after fine-tuning \texttt{Qwen3-1.7B} on \textbf{4K successful trajectories} sampled from \methodname-14B with the training loss approaching \emph{almost zero} (Figure~\ref{fig:codescout_1.7B_RFT}), we find that training this checkpoint further using reinforcement learning improves the model performance even further - we observe an increasing reward curve with a steep slope during the first $\approx$ 20 training steps before saturating. 

\input{Tables/baselines_tool_descriptions}
\section{Additional Discussion of Related Work}\label{appendix:related_work_detailed}
\input{Sections/7_related_work}

\section{Limitations and Directions for Future Work}\label{appendix:future_work}
Although the proposed \methodname recipe is significantly more general and scalable than most prior approaches to code localization, several important limitations remain. First, our training and evaluation experiments are largely restricted to Python repositories. This constraint arises primarily from the lack of large-scale training counterparts to evaluation benchmarks such as Multi-SWE-Bench~\citep{zan2025multiswebench}, which contains issues from repositories written in languages other than Python. Moreover, extracting ground-truth localization targets requires programming language-specific processing of ground-truth code patches, which not only limits our ability to train agents for other programming languages but also evaluate zero-shot transfer of our models to these languages. However, note that these are not necessarily limitations of our agent scaffold or RL recipe and we leave it for future work to curate training datasets and develop patch processing methods (for extracting ground truth locations) for a broader range of programming languages.

Another limitation is that our ground-truth localization targets are primarily derived from the modified regions of the reference code patches. While this provides a practical approximation of the relevant files, classes, and functions, it may omit other code segments that are important for understanding or resolving an issue but do not require direct modification. Note that this limitation also exists in all the baselines we compare \methodname against and is non-trivial to address because determining relevance of a code location (file/module/class) to a given GitHub issue is subjective and curating this data at a training scale makes this problem even harder to address.

Finally, localizing relevant/faulty code from the repository given an issue description does not cover the broader scope of repository-level code search tasks, and our work does not address general-purpose repository-level question-answering/search tasks like those covered by benchmarks like CodeAssistBench~\citep{kim2026codeassistbenchcabdataset} and SWE-QA~\citep{peng2025sweqalanguagemodelsanswer}. Future research directions include expanding the scope of our work to a broader range of tasks requiring repository-level code localization.

\section{How does choice of RL algorithm impact localization performance?}\label{appendix:rl_ablation_study}

In \S\ref{sec:train_algo}, we described our choice of RL algorithm (GSPO) with modifications like no advantage standardization. A natural question is whether this particular configuration is optimal, or whether better performance can be achieved through other RL algorithms. We investigate this by training \texttt{Qwen3-4B-Instruct-2507} under a range of algorithm configurations, all sharing the same reward function which augments the reward from Equation~\ref{eqn:reward} with the auxiliary binary term $r^{\mathsf{turn}}$.

\emph{Comparability note.} These ablation runs share the same learning rate, batch size, and number of rollouts as \methodname-4B (\S\ref{sec:train_setup}), but use 4 turns (vs.\ 6) in its system prompt and the auxiliary turn reward $r^{\mathsf{turn}}(\tau, k)$ with $k$ = 4. All ablation runs are trained for 200 steps; and small performance differences may fall within the range of training variance. We compare all the checkpoints by evaluating them on SWE-Bench Pro using the setup from \S\ref{sec:eval_benchmarks}.

\paragraph{Recipe-level comparison.}
We first compare four representative critic-free policy-gradient recipes. These methods avoid the memory overhead of a separate value network, making them practical for multi-turn agentic training with long trajectories. They differ primarily in the \emph{policy loss type} (how the importance ratio is computed and constrained) and the \emph{loss reduction} (how per-token losses are aggregated):
\begin{itemize}
    \item \textbf{GSPO}~\citep{zheng2025groupsequencepolicyoptimization} (our configuration): sequence-level importance ratio with tighter clipping ($\epsilon = $ 3e-4/4e-4), sequence-mean loss reduction, and no advantage standardization.
    \item \textbf{GRPO}~\citep{shao2024deepseekmathpushinglimitsmathematical}: token-level PPO-clip loss ($\epsilon = $ 0.2) with advantage standardization.
    \item \textbf{SAPO}~\citep{gao2025softadaptivepolicyoptimization}: replaces PPO clipping with a soft gating function that applies asymmetric temperatures to positive- and negative-advantage tokens, with advantage standardization.
    \item \textbf{Dr.GRPO}~\citep{liu2025understandingr1zeroliketrainingcritical}: token-level PPO-clip loss ($\epsilon$=0.2) with length-unbiased loss reduction, without advantage standardization.
\end{itemize}
As shown in the top group of Table~\ref{tab:results_ablation}, all four recipes achieve file-level F1 within a range of 47--55\% and function-level F1 within the range of 22--25\%, suggesting that our task is somewhat insensitive to the choice of RL algorithm. GSPO achieves the highest file-level F1, while Dr.GRPO leads on module- and function-level F1 score.

\paragraph{Single-factor ablations.}
We next isolate the effect of individual design choices by changing exactly one factor from the GSPO configuration.

\emph{Loss reduction.} Standard GRPO normalizes the per-sequence loss by the response length, which introduces a length bias~\citep{liu2025understandingr1zeroliketrainingcritical}. Dr.GRPO addresses this by dividing the per-sequence token sum by a fixed constant (the generation budget) instead. Inspired by the adoption of this technique in recent work~\citep{cognition2025swegrep}, we test whether it benefits GSPO by switching from \texttt{sequence\_mean} to \texttt{seq\_mean\_token\_sum\_norm} while keeping all other settings unchanged. This causes a substantial drop in file-level F1 (54.83\% $\to$ 42.02\%), suggesting it does not transfer well to GSPO's sequence-level importance ratio.

\emph{Advantage normalization.} GRPO and SAPO both standardize advantages by their standard deviation, while GSPO does not. Enabling this by setting \texttt{grpo\_norm\_by\_std=true} slightly lowers file-level F1 (54.83\% $\to$ 52.73\%) but improves module- and function-level metrics (31.43\% $\to$ \textbf{34.37\%} and 23.29\% $\to$ \textbf{26.41\%}).

These findings indicate that the GSPO configuration used by \methodname performs competitively across all granularities, supporting its use as a reasonable default. More broadly, the relatively narrow performance range across recipes indicates that, for the code localization task with \texttt{Qwen3-4B/14B}, the choice of reward design (\S\ref{sec:reward_design}) and agent scaffold (\S\ref{sec:agent_scaffold}) likely matters more than the specific RL algorithm, consistent with our recipe-oriented perspective.

\input{Tables/results_train_ablation_detailed}

\section{Example Trajectories for \methodname-4B and \methodname-14B}\label{appendix:traj_examples}
This section provides some example trajectories sampled from \methodname-4B and \methodname-14B models using our evaluation setup (\S\ref{sec:eval_benchmarks}). We use an identical GitHub issue (\texttt{django\_\_django-13363}) from the SWE-Bench Verified benchmark~\citep{chowdhury2024swebenchverified} for both these models to allow a direct comparison of their tool-use behaviors and problem-solving approaches. Note that both models achieve a perfectly correct localization for this instance, i.e. they achieve an F1 score of 1.0 for all three granularities: file, module, and function. Note that we omit the system prompt and user prompt (which mentions the issue description) for these example rollouts. Figures \ref{fig:traj_14B_p1} and \ref{fig:traj_14b_p2} present the trajectory sampled from \methodname-14B, and figures \ref{fig:traj_4b_p2} and \ref{fig:traj_4b_p2} present the trajectory sampled from \methodname-4B.

As noted in \S\ref{sec:tool_use_behaviour}, \methodname-14B relies only on two Unix command-line utilities for code localization: \texttt{ripgrep (rg)} and \texttt{sed}. By analyzing the trajectory, we observe that the model uses \texttt{ripgrep} for efficient keyword-based search of the code repository and \texttt{sed} for reading specific line-ranges from relevant files. Therefore, by exploring the repository for \emph{only} 3 steps and using \emph{only} 2 command-line utilities, \methodname-14B can determine the exact set of relevant set of files, classes, and functions that require modification to fix a given GitHub issue, highlighting the effectiveness of our approach.

Interestingly, \methodname-4B uses a few more Unix command-line utilities for code localization: \texttt{rg}, \texttt{find}, \texttt{cat}, \texttt{xargs} and \texttt{sed}. The model mostly uses \texttt{rg}, \texttt{find} and \texttt{xargs} during keyword-based searches, and \texttt{sed} and \texttt{cat} for reading particular line-ranges from files. Furthermore, the bash commands issued by \methodname-4B are more complex and often use more than one Unix utility in the same command chained together through piping ($|$). Here as well, \methodname-4B performs code localization by exploring the repository for \emph{only} 3 steps and using \emph{only} 5 command-line utilities. For both the models, the command-line utilities used in these examples align with those used by the final training checkpoint in Figure~\ref{fig:merged_tool_use}.

A noticeable characteristic of \methodname-4B is that it frequently performs parallel tool-calling unlike \methodname-14B. We hypothesize that this is because of two reasons: (1) our training algorithm, and (2) choice of base model for these models. Firstly, our training algorithm does not reward/encourage parallel tool-calling and this is mainly mentioned through the system prompt: thus rollouts with parallel tool-calling have no advantage over those without any parallel tool-calling during training. Secondly, the base model for \methodname-4B, \texttt{Qwen3-4B-Instruct-2507}, has strong instruction-following capabilities (stronger than \texttt{Qwen3-4B}, and possibly even better than \texttt{Qwen3-14B} which is the base model for \methodname-14B). As a result, the simpler prompting mechanism is sufficient for our 4B model but not for the 14B model.

\section{Additional Comparison with Scaffolds that Predict a Ranked List of Top-$K$ Locations}\label{appendix:topk_discussion}
As opposed to prior approaches that a fixed number of top-$K$ relevant locations RepoNavigator~\citep{zhang2025one} and \methodname are flexible and allow the agent to dynamically predict a variable number of locations. This has the added advantage of not having to choose an appropriate value of $K$ based on the domain or choice of evaluation benchmark. In particular, for our choice of evaluation benchmarks, the design choice of predicting a ranked list of top-$K$ locations (with $K$ generally set to 5) results in a higher recall but lower precision. Moreover, this design is particularly not suitable for RL fine-tuning of LLM agents, especially when $K$ is larger than the number of ground truth locations as there is no way to determine the relevance of the remaining locations predicted by the agent and the reward function will generally ignore these predictions completely during training.

Note that the choice of $K$ will also determine the trade-off between precision and recall and the results in \S\ref{sec:results} for these baselines are based on the corresponding default choice of $K$ used by their scaffolds. We also include comparisons with these methods for different choices of $K$ to understand the precision-recall tradeoff and to compare these methods with \methodname-4B and \methodname-14B models. Table~\ref{tab:return_at_k_swe_bench_verified} presents the detailed results for three choices of $K$ - 1, 3, and all (the default setting where K is set to 5). Impressively, \methodname models outperform all prior methods that use a larger LLM across different choices of K even when dynamically predicting a variable number of locations. Furthermore, note that there is significant variance in the precision, recall, and F1 scores of the baseline methods for different choices of K implying that deciding the right value of K for the optimal performance is non-trivial and also depends on the characteristics of the GitHub issue and dataset. Impressively, \methodname strikes the right balance between precision and recall as a result of its scaffold design and our RL recipe that incorporates both precision and recall by computing the F1 scores.
\begin{table*}[!t]
\caption{Comparison of \methodname with methods that predict a ranked list of top-K locations for different choices of K (Return@K) on \textbf{SWE-Bench Verified}. The setting \texttt{all} refers to the default setting with K$=$5 used by these scaffolds.}
\centering
\resizebox{\textwidth}{!}{%
\begin{tabular}{lc|ccc|ccc}
\toprule
\multirow{2}{*}{\textbf{Scaffold + LLM}} & \multirow{2}{*}{\textbf{Return@K}} & \multicolumn{3}{c|}{\textbf{File}} & \multicolumn{3}{c}{\textbf{Function}} \\
& & F1 & Prec. & Rec. & F1 & Prec. & Rec. \\
\midrule
\multirow{3}{*}{CoSIL + Qwen2.5-32B-Instruct} & 1 & 57.25 & 64.65 & 51.36 & 29.11 & 42.51 & 22.13 \\
& 3 & 41.60 & 29.53 & 70.30 & 25.35 & 20.80 & 32.46 \\
& all & 30.77 & 19.34 & 83.50 & 22.11 & 14.85 & 55.38 \\
\midrule
\multirow{3}{*}{Agentless + Qwen2.5-32B-Instruct} & 1 & 56.17 & 63.93 & 50.08 & 22.18 & 32.78 & 16.75 \\
& 3 & 41.46 & 30.13 & 66.45 & 23.51 & 20.83 & 26.98 \\
& all & 35.38 & 25.60 & 78.93 & 27.33 & 24.07 & 40.97 \\
\midrule
\multirow{3}{*}{LocAgent + Qwen2.5-32B-Instruct} & 1 & 64.66 & 78.31 & 55.06 & 32.78 & 52.06 & 23.92 \\
& 3 & 49.34 & 38.54 & 68.54 & 27.87 & 24.60 & 32.14 \\
& all & 44.18 & 34.18 & 79.39 & 21.48 & 16.29 & 46.79 \\
\midrule
\methodname-4B & - & \underline{68.52} & 71.53 & 67.74 & \underline{36.78} & 40.71 & 35.72 \\
\midrule
\methodname-14B & - & \textbf{68.57} & 71.00 & 68.69 & \textbf{40.32} & 43.74 & 40.27 \\
\bottomrule
\end{tabular}%
}
\label{tab:return_at_k_swe_bench_verified}
\end{table*}
\begin{figure*}[!t]
    \centering
\begin{tcolorbox}[colback=blue!3!white, colframe=blue!75!black, title={\small Action 1 (Terminal Tool)}]
\begin{lstlisting}[basicstyle=\fontsize{7.5}{8.5}\ttfamily, breaklines=true]
rg "TruncDate" -t py
\end{lstlisting}
\end{tcolorbox}

\begin{tcolorbox}[colback=orange!5!white, colframe=orange!75!black, title={\small Observation 1}]
\begin{lstlisting}[basicstyle=\fontsize{7.5}{8.5}\ttfamily, breaklines=true]
django/db/models/functions/__init__.py
5:    ExtractWeekDay, ExtractYear, Now, Trunc, TruncDate, TruncDay, TruncHour,
30:    'TruncDate', 'TruncDay', 'TruncHour', 'TruncMinute', 'TruncMonth',

django/db/models/functions/datetime.py
287:class TruncDate(TruncBase):
325:DateTimeField.register_lookup(TruncDate)

tests/db_functions/datetime/test_extract_trunc.py
13:    ExtractWeekDay, ExtractYear, Trunc, TruncDate, TruncDay, TruncHour,
839:            DTModel.objects.annotate(extracted=TruncDate('start_datetime')).order_by('start_datetime'),
\end{lstlisting}
\end{tcolorbox}

\begin{tcolorbox}[colback=blue!3!white, colframe=blue!75!black, title={\small Action 2 (Terminal Tool)}]
\begin{lstlisting}[basicstyle=\fontsize{7.5}{8.5}\ttfamily, breaklines=true]
sed -n '1,500p' django/db/models/functions/datetime.py
\end{lstlisting}
\end{tcolorbox}

\begin{tcolorbox}[colback=orange!5!white, colframe=orange!75!black, title={\small Observation 2 (Trimmed)}]
\begin{lstlisting}[basicstyle=\fontsize{7.5}{8.5}\ttfamily, breaklines=true]
from django.db.models.fields import (
    DateField, DateTimeField, DurationField, Field, IntegerField, TimeField,
)
... 

class TruncBase(TimezoneMixin, Transform):
    def as_sql(self, compiler, connection):
        inner_sql, inner_params = compiler.compile(self.lhs)
        if isinstance(self.output_field, DateTimeField):
            tzname = self.get_tzname()
            sql = connection.ops.datetime_trunc_sql(self.kind, inner_sql, tzname)
...

class TruncDate(TruncBase):
    kind = 'date'
    lookup_name = 'date'
    output_field = DateField()

    def as_sql(self, compiler, connection):
        # Cast to date rather than truncate to date.
        lhs, lhs_params = compiler.compile(self.lhs)
        tzname = timezone.get_current_timezone_name() if settings.USE_TZ else None
        sql = connection.ops.datetime_cast_date_sql(lhs, tzname)
        return sql, lhs_params

class TruncTime(TruncBase):
    kind = 'time'
    lookup_name = 'time'
    output_field = TimeField()
...
\end{lstlisting}
\end{tcolorbox}
\caption{Example trajectory sampled from \methodname-14B for the task \texttt{django\_\_django-13363} (continued on the next page).}
\label{fig:traj_14B_p1}
\end{figure*}
\begin{figure*}[!ht]

\begin{tcolorbox}[colback=blue!3!white, colframe=blue!75!black, title={\small Action 3 (Terminal Tool)}]
\begin{lstlisting}[basicstyle=\fontsize{7.5}{8.5}\ttfamily, breaklines=true]
rg "get_current_timezone_name" -t py
\end{lstlisting}
\end{tcolorbox}

\begin{tcolorbox}[colback=orange!5!white, colframe=orange!75!black, title={\small Observation 3}]
\begin{lstlisting}[basicstyle=\fontsize{7.5}{8.5}\ttfamily, breaklines=true]
django/db/models/functions/datetime.py
25:                tzname = timezone.get_current_timezone_name()
295:        tzname = timezone.get_current_timezone_name() if settings.USE_TZ else None
308:        tzname = timezone.get_current_timezone_name() if settings.USE_TZ else None

django/utils/timezone.py
64:def get_current_timezone_name():
\end{lstlisting}
\end{tcolorbox}

\begin{tcolorbox}[colback=blue!3!white, colframe=blue!75!black, title={\small Action 4 (Localization Finish Tool - Agent Loop Terminated)}]
\begin{lstlisting}[basicstyle=\fontsize{7.5}{8.5}\ttfamily, breaklines=true]
[
    {
        "file": "django/db/models/functions/datetime.py",
        "class_name": "TruncDate",
        "function_name": "as_sql"
    },
    {
        "file": "django/db/models/functions/datetime.py",
        "class_name": "TruncTime",
        "function_name": "as_sql"
    }
]
\end{lstlisting}
\end{tcolorbox}

\begin{tcolorbox}[colback=violet!3!white, colframe=violet!75!black, title={\small Reward Computation}]
\begin{lstlisting}[basicstyle=\fontsize{7.5}{8.5}\ttfamily, breaklines=true]
Ground truth locations:
    Files: django/db/models/functions/datetime.py
    Modules: django/db/models/functions/datetime.py:TruncDate, django/db/models/functions/datetime.py:TruncTime
    Functions: django/db/models/functions/datetime.py:TruncDate.as_sql, django/db/models/functions/datetime.py:TruncTime.as_sql
Rewards:
    File-level F1 score: 1.0
    Module-level F1 score: 1.0
    Function-level F1 score: 1.0
\end{lstlisting}
\end{tcolorbox}
\caption{Example trajectory sampled from \methodname-14B for the task \texttt{django\_\_django-13363}.}
\label{fig:traj_14b_p2}
\end{figure*}

\begin{figure*}[!ht]
    \centering

\begin{tcolorbox}[colback=blue!3!white, colframe=blue!75!black, title={\small Action 1 (Terminal Tool - 4 parallel calls)}]
\begin{lstlisting}[basicstyle=\fontsize{7.5}{8.5}\ttfamily, breaklines=true]
# Command 1
find . -type f -name "*.py" | xargs rg -l TruncDate
# Command 2
find . -type f -name "*.py" | xargs rg -l TruncTime
# Command 3
rg -l TimeZone
# Command 4
rg -l TruncBase
\end{lstlisting}
\end{tcolorbox}

\begin{tcolorbox}[colback=orange!5!white, colframe=orange!75!black, title={\small Observation 1 (trimmed)}]
\begin{lstlisting}[basicstyle=\fontsize{7.5}{8.5}\ttfamily, breaklines=true]
# [Results for Command 1]
./django/db/models/functions/datetime.py
./django/db/models/functions/__init__.py
./tests/db_functions/datetime/test_extract_trunc.py

# [Results for Command 2]
./django/db/models/functions/datetime.py
./django/db/models/functions/__init__.py
./tests/db_functions/datetime/test_extract_trunc.py

# [Results for Command 3]
django/db/backends/postgresql/base.py
django/templatetags/tz.py
tests/db_functions/datetime/test_extract_trunc.py
... (truncated)

# [Results for Command 4]
django/db/models/functions/datetime.py
\end{lstlisting}
\end{tcolorbox}
\begin{tcolorbox}[colback=blue!3!white, colframe=blue!75!black, title={\small Action 2 (Terminal Tool - 3 parallel calls)}]
\begin{lstlisting}[basicstyle=\fontsize{7.5}{8.5}\ttfamily, breaklines=true]
# Command 1
cat django/db/models/functions/datetime.py | sed -n '1,500p'
# Command 2
cat django/db/models/functions/datetime.py | sed -n '501,1000p'
# Command 3
rg -n 'get_current_timezone_name' django/db/models/functions/datetime.py
\end{lstlisting}
\end{tcolorbox}

\begin{tcolorbox}[colback=orange!5!white, colframe=orange!75!black, title={\small Observation 2 (trimmed)}]
\begin{lstlisting}[basicstyle=\fontsize{7.5}{8.5}\ttfamily, breaklines=true]
# [Results for Command 1]
...
from django.utils import timezone
class TimezoneMixin:
    def get_tzname(self):
...
class TruncDate(TruncBase):
    def as_sql(self, compiler, connection):
        tzname = timezone.get_current_timezone_name() if settings.USE_TZ else None
        sql = connection.ops.datetime_cast_date_sql(lhs, tzname)
...

# [Results for Command 2 (empty)]

# [Results for Command 3]
25:                tzname = timezone.get_current_timezone_name()
295:        tzname = timezone.get_current_timezone_name() if settings.USE_TZ else None
308:        tzname = timezone.get_current_timezone_name() if settings.USE_TZ else None
\end{lstlisting}
\end{tcolorbox}
\caption{Example trajectory sampled from \methodname-4B for the task \texttt{django\_\_django-13363} (continued on the next page).}
\label{fig:traj_4b_p1}
\end{figure*}

\begin{figure*}[!ht]
    \centering
\begin{tcolorbox}[colback=blue!3!white, colframe=blue!75!black, title={\small Action 3 (Terminal Tool - 2 parallel calls)}]
\begin{lstlisting}[basicstyle=\fontsize{7.5}{8.5}\ttfamily, breaklines=true]
# Command 1
rg -n 'datetime_cast_date_sql' django/db/models/functions/datetime.py
# Command 2
rg -n 'datetime_cast_time_sql' django/db/models/functions/datetime.py
\end{lstlisting}
\end{tcolorbox}

\begin{tcolorbox}[colback=orange!5!white, colframe=orange!75!black, title={\small Observation 3}]
\begin{lstlisting}[basicstyle=\fontsize{7.5}{8.5}\ttfamily, breaklines=true]
# [Results for Command 1]
296:        sql = connection.ops.datetime_cast_date_sql(lhs, tzname)

# [Results for Command 2]
309:        sql = connection.ops.datetime_cast_time_sql(lhs, tzname)
\end{lstlisting}
\end{tcolorbox}

\begin{tcolorbox}[colback=blue!3!white, colframe=blue!75!black, title={\small Action 4 (Localization Finish Tool - Agent Loop Terminated)}]
\begin{lstlisting}[basicstyle=\fontsize{7.5}{8.5}\ttfamily, breaklines=true]
[
    {
        "file": "django/db/models/functions/datetime.py",
        "class_name": "TruncDate",
        "function_name": "as_sql"
    },
    {
        "file": "django/db/models/functions/datetime.py",
        "class_name": "TruncTime",
        "function_name": "as_sql"
    }
]
\end{lstlisting}
\end{tcolorbox}
\begin{tcolorbox}[colback=violet!3!white, colframe=violet!75!black, title={\small Reward Computation}]
\begin{lstlisting}[basicstyle=\fontsize{7.5}{8.5}\ttfamily, breaklines=true]
Ground truth locations:
    Files: django/db/models/functions/datetime.py
    Modules: django/db/models/functions/datetime.py:TruncDate, django/db/models/functions/datetime.py:TruncTime
    Functions: django/db/models/functions/datetime.py:TruncDate.as_sql, django/db/models/functions/datetime.py:TruncTime.as_sql
Rewards:
    File-level F1 score: 1.0
    Module-level F1 score: 1.0
    Function-level F1 score: 1.0
\end{lstlisting}
\end{tcolorbox}
\caption{Example trajectory sampled from \methodname-4B for the task \texttt{django\_\_django-13363}.}
\label{fig:traj_4b_p2}
\end{figure*}
\section{Prompts for Issue Resolution with Localization-Augmented Context}\label{appendix:prompts_swe_bench}
We provide detailed user and system prompts in the OpenHands software agent SDK across various experimental settings used to demonstrate the advantages of augmenting issue resolution agents with relevant localization context (\S\ref{sec:downstream_swebench}). For the vanilla setup that is not augmented with any additional localization context, we provide the user prompt in Figure~\ref{fig:user_prompt_vanilla}. Figures~\ref{fig:user_prompt_codescout_augment1} and~\ref{fig:user_prompt_codescout_augment2} describe the user prompt for the experimental setting where the issue resolution agent is augmented with locations retrieved by \methodname-14B. Figures~\ref{fig:user_prompt_oracle_augment1} and~\ref{fig:user_prompt_oracle_augment2} describe the user prompt for the experimental setting where the issue resolution agent is augmented with oracle locations. All three settings use the default system prompt for OpenHands Software Agent SDK which is presented in Figures ~\ref{fig:swe_bench_system_prompt1},~\ref{fig:swe_bench_system_prompt2},~\ref{fig:swe_bench_system_prompt3},~\ref{fig:swe_bench_system_prompt4}, and~\ref{fig:swe_bench_system_prompt5}. 

\begin{figure*}[!h]
    \centering
\begin{tcolorbox}[colback=purple!6!white, colframe=purple!75!black, title={\small User Prompt for OpenHands Issue Resolution Agent (Vanilla Setting)}]
\begin{lstlisting}[basicstyle=\fontsize{7.7}{8}\ttfamily]
I have access to a python code repository in the directory {{ instance.repo_path }} . You can explore and modify files using the available tools. Consider the following issue description:

<issue_description>
{{ instance.problem_statement }}
</issue_description>

Can you help me implement the necessary changes to the repository so that the requirements specified in the <issue_description> are met?
I've already taken care of all changes to any of the test files described in the <issue_description>. This means you DON'T have to modify the testing logic or any of the tests in any way!
Also the development Python environment is already set up for you (i.e., all dependencies already installed), so you don't need to install other packages.
Your task is to make the minimal changes to non-test files in the {{ instance.repo_path }} directory to ensure the <issue_description> is satisfied.

Follow these phases to resolve the issue:

Phase 1. READING: read the problem and reword it in clearer terms
   1.1 If there are code or config snippets. Express in words any best practices or conventions in them.
   1.2 Hightlight message errors, method names, variables, file names, stack traces, and technical details.
   1.3 Explain the problem in clear terms.
   1.4 Enumerate the steps to reproduce the problem.
   1.5 Hightlight any best practices to take into account when testing and fixing the issue

Phase 2. RUNNING: install and run the tests on the repository
   2.1 Activate the environment by running  
       ./opt/miniconda3/etc/profile.d/conda.sh ; conda activate testbed
   2.2 Follow the readme
   2.3 Install the environment and anything needed
   2.4 Iterate and figure out how to run the tests

Phase 3. EXPLORATION: find the files that are related to the problem and possible solutions
   3.1 Use `grep` to search for relevant methods, classes, keywords and error messages.
   3.2 Identify all files related to the problem statement.
   3.3 Propose the methods and files to fix the issue and explain why.
   3.4 From the possible file locations, select the most likely location to fix the issue.

Phase 4. TEST CREATION: before implementing any fix, create a script to reproduce and verify the issue.
   4.1 Look at existing test files in the repository to understand the test format/structure.
   4.2 Create a minimal reproduction script that reproduces the located issue.
   4.3 Run the reproduction script to confirm you are reproducing the issue.
   4.4 Adjust the reproduction script as necessary.

Phase 5. FIX ANALYSIS: state clearly the problem and how to fix it
   5.1 State clearly what the problem is.
   5.2 State clearly where the problem is located.
   5.3 State clearly how the test reproduces the issue.
   5.4 State clearly the best practices to take into account in the fix.
   5.5 State clearly how to fix the problem.

Phase 6. FIX IMPLEMENTATION: Edit the source code to implement your chosen solution.
   6.1 Make minimal, focused changes to fix the issue.

Phase 7. VERIFICATION: Test your implementation thoroughly.
   7.1 Run your reproduction script to verify the fix works.
   7.2 Add edge cases to your test script to ensure comprehensive coverage.
   7.3 Run existing tests related to the modified code to ensure you haven't broken anything.

8. FINAL REVIEW: Carefully re-read the problem description and compare your changes with the base commit {{ instance.base_commit }}.
   8.1 Ensure you've fully addressed all requirements.
   8.2 Run any tests in the repository related to:
     8.2.1 The issue you are fixing
     8.2.2 The files you modified
     8.2.3 The functions you changed
   8.3 If any tests fail, revise your implementation until all tests pass

Be thorough in your exploration, testing, and reasoning. It's fine if your thinking process is lengthy - quality and completeness are more important than brevity.

\end{lstlisting}
\end{tcolorbox}
\caption{User prompt for OpenHands issue resolution agent in vanilla setup without augmenting it with any localization results (\S\ref{sec:downstream_swebench}).}
\label{fig:user_prompt_vanilla}
\end{figure*}

\begin{figure*}[!h]
    \centering
\begin{tcolorbox}[colback=blue!6!white, colframe=blue!45!white, title={\small User Prompt for OpenHands Issue Resolution Agent Augmented with Locations Retrieved by \methodname-14B (continued)}]
\begin{lstlisting}[basicstyle=\fontsize{7.7}{8}\ttfamily]
I have access to a python code repository in the directory {{ instance.repo_path }} . You can explore and modify files using the available tools. Consider the following issue description:

<issue_description>
{{ instance.problem_statement }}
</issue_description>

A separate search subagent has already explored the repository to identify files, modules, and functions that are likely relevant to the issue.

<relevant_context>
{{ instance.search_results }}
</relevant_context>

Important notes about this context:
- The relevant context is intended to accelerate navigation, not replace your own judgment.
- The listed files/functions are very likely but not guaranteed to be sufficient.
- You may inspect files outside this list ONLY if needed, but you should prioritize the provided context.
- If you determine that some listed items are irrelevant or that additional files are required, state this explicitly in your analysis.

Can you help me implement the necessary changes to the repository so that the requirements specified in the <issue_description> are met?
I've already taken care of all changes to any of the test files described in the <issue_description>. This means you DON'T have to modify the testing logic or any of the tests in any way!
Also the development Python environment is already set up for you (i.e., all dependencies already installed), so you don't need to install other packages.
Your task is to make the minimal changes to non-test files in the {{ instance.repo_path }} directory to ensure the <issue_description> is satisfied.

Follow these phases to resolve the issue:

Phase 1. READING: read the problem and reword it in clearer terms
   1.1 If there are code or config snippets. Express in words any best practices or conventions in them.
   1.2 Hightlight message errors, method names, variables, file names, stack traces, and technical details.
   1.3 Explain the problem in clear terms.
   1.4 Enumerate the steps to reproduce the problem.
   1.5 Hightlight any best practices to take into account when testing and fixing the issue

Phase 2. RUNNING: install and run the tests on the repository
   2.1 Activate the environment by running  
       ./opt/miniconda3/etc/profile.d/conda.sh ; conda activate testbed
   2.2 Follow the readme
   2.3 Install the environment and anything needed
   2.4 Iterate and figure out how to run the tests

Phase 3. CONTEXT REVIEW & CODE EXPLORATION: use the relevant context to orient yourself in the codebase.
   3.1 Review the <relevant_context> and understand why each listed file or function may be relevant.
   3.2 Prioritize and inspect the files and functions mentioned in the <relevant_context>.
   3.3 Read the relevant source code to understand its behavior.
   3.4 Identify discrepancies between current behavior and the expectations in the <issue_description>.
   3.5 Document any irrelevant items or missing files and adjust exploration as needed.

Phase 4. TEST CREATION: before implementing any fix, create a script to reproduce and verify the issue.
   4.1 Look at existing test files in the repository to understand the test format/structure.
   4.2 Create a minimal reproduction script that reproduces the located issue.
   4.3 Run the reproduction script to confirm you are reproducing the issue.
   4.4 Adjust the reproduction script as necessary.

Phase 5. FIX ANALYSIS: state clearly the problem and how to fix it
   5.1 State clearly what the problem is.
   5.2 State clearly where the problem is located.
   5.3 State clearly how the test reproduces the issue.
   5.4 State clearly the best practices to take into account in the fix.
   5.5 State clearly how to fix the problem.
\end{lstlisting}
\end{tcolorbox}
\caption{User prompt for OpenHands issue resolution agent augmented with locations retrieved by \methodname-14B (\S\ref{sec:downstream_swebench}).}
\label{fig:user_prompt_codescout_augment1}
\end{figure*}
\begin{figure*}[!ht]
    \centering
\begin{tcolorbox}[colback=blue!6!white, colframe=blue!45!white, title={\small User Prompt for OpenHands Issue Resolution Agent Augmented with Locations Retrieved by \methodname-14B (continued)}]
\begin{lstlisting}[basicstyle=\fontsize{7.7}{8}\ttfamily]
Phase 6. FIX IMPLEMENTATION: Edit the source code to implement your chosen solution.
   6.1 Make minimal, focused changes to fix the issue.

Phase 7. VERIFICATION: Test your implementation thoroughly.
   7.1 Run your reproduction script to verify the fix works.
   7.2 Add edge cases to your test script to ensure comprehensive coverage.
   7.3 Run existing tests related to the modified code to ensure you haven't broken anything.

Phase 8. FINAL REVIEW: Carefully re-read the problem description and compare your changes with the base commit {{ instance.base_commit }}.
   8.1 Ensure you've fully addressed all requirements.
   8.2 Run any tests in the repository related to:
     8.2.1 The issue you are fixing
     8.2.2 The files you modified
     8.2.3 The functions you changed
   8.3 If any tests fail, revise your implementation until all tests pass

Be thorough in your exploration, testing, and reasoning. It's fine if your thinking process is lengthy - quality and completeness are more important than brevity.
\end{lstlisting}
\end{tcolorbox}
\caption{User prompt for OpenHands issue resolution agent augmented with locations retrieved by \methodname-14B (\S\ref{sec:downstream_swebench}).
\label{fig:user_prompt_codescout_augment2}
}
\end{figure*}

\begin{figure*}[!h]
    \centering
\begin{tcolorbox}[colback=red!6!white, colframe=red!75!black, title={\small User Prompt for OpenHands Issue Resolution Agent Augmented with Oracle Locations (continued)}]
\begin{lstlisting}[basicstyle=\fontsize{7.7}{8}\ttfamily]
I have access to a python code repository in the directory {{ instance.repo_path }} . You can explore and modify files using the available tools. Consider the following issue description:

<issue_description>
{{ instance.problem_statement }}
</issue_description>

An oracle has already determined the exact set of files, modules and functions that need modification to resolve the above issue description.

<oracle_context>
{{ instance.search_results }}
</oracle_context>

Important notes about this context:
- The oracle context is guaranteed to be exhaustive and precise for resolving the issue.
- In cases where the context doesn't mention functions or modules, it means that the changes either belong to the global scope of the file or involve adding new functions/modules in the given file.
- You may inspect files outside this list ONLY if needed, but you primarily should focus on the provided context as that is the one which requires modification to resolve the issue.
- The provided context will not include files that need to be created from scratch for resolving the issue.

Can you help me implement the necessary changes to the repository so that the requirements specified in the <issue_description> are met?
I've already taken care of all changes to any of the test files described in the <issue_description>. This means you DON'T have to modify the testing logic or any of the tests in any way!
Also the development Python environment is already set up for you (i.e., all dependencies already installed), so you don't need to install other packages.
Your task is to make the minimal changes to non-test files in the {{ instance.repo_path }} directory to ensure the <issue_description> is satisfied.
\end{lstlisting}
\end{tcolorbox}
\caption{User prompt for OpenHands issue resolution agent augmented with oracle locations (\S\ref{sec:downstream_swebench}).
\label{fig:user_prompt_oracle_augment1}
}
\end{figure*}

\begin{figure*}[!h]
    \centering
\begin{tcolorbox}[colback=red!6!white, colframe=red!75!black, title={\small User Prompt for OpenHands Issue Resolution Agent Augmented with Oracle Locations}]
\begin{lstlisting}[basicstyle=\fontsize{7.7}{8}\ttfamily]
Follow these phases to resolve the issue:

Phase 1. READING: read the problem and reword it in clearer terms
   1.1 If there are code or config snippets. Express in words any best practices or conventions in them.
   1.2 Hightlight message errors, method names, variables, file names, stack traces, and technical details.
   1.3 Explain the problem in clear terms.
   1.4 Enumerate the steps to reproduce the problem.
   1.5 Hightlight any best practices to take into account when testing and fixing the issue

Phase 2. RUNNING: install and run the tests on the repository
   2.1 Activate the environment by running  
       ./opt/miniconda3/etc/profile.d/conda.sh ; conda activate testbed
   2.2 Follow the readme
   2.3 Install the environment and anything needed
   2.4 Iterate and figure out how to run the tests

Phase 3. CONTEXT REVIEW & CODE EXPLORATION: use the oracle context to orient yourself in the codebase.
   3.1 Review the <oracle_context> and understand why each listed file or function needs modification.
   3.2 Prioritize and inspect the files and functions mentioned in the <oracle_context>.
   3.3 Read the relevant source code to understand its behavior.
   3.4 Identify discrepancies between current behavior and the expectations in the <issue_description>.

Phase 4. TEST CREATION: before implementing any fix, create a script to reproduce and verify the issue.
   4.1 Look at existing test files in the repository to understand the test format/structure.
   4.2 Create a minimal reproduction script that reproduces the located issue.
   4.3 Run the reproduction script to confirm you are reproducing the issue.
   4.4 Adjust the reproduction script as necessary.

Phase 5. FIX ANALYSIS: state clearly the problem and how to fix it
   5.1 State clearly what the problem is.
   5.2 State clearly where the problem is located.
   5.3 State clearly how the test reproduces the issue.
   5.4 State clearly the best practices to take into account in the fix.
   5.5 State clearly how to fix the problem.

Phase 6. FIX IMPLEMENTATION: Edit the source code to implement your chosen solution.
   6.1 Make minimal, focused changes to fix the issue.

Phase 7. VERIFICATION: Test your implementation thoroughly.
   7.1 Run your reproduction script to verify the fix works.
   7.2 Add edge cases to your test script to ensure comprehensive coverage.
   7.3 Run existing tests related to the modified code to ensure you haven't broken anything.

Phase 8. FINAL REVIEW: Carefully re-read the problem description and compare your changes with the base commit {{ instance.base_commit }}.
   8.1 Ensure you've fully addressed all requirements.
   8.2 Run any tests in the repository related to:
     8.2.1 The issue you are fixing
     8.2.2 The files you modified
     8.2.3 The functions you changed
   8.3 If any tests fail, revise your implementation until all tests pass

Be thorough in your exploration, testing, and reasoning. It's fine if your thinking process is lengthy - quality and completeness are more important than brevity.\end{lstlisting}
\end{tcolorbox}
\caption{User prompt for OpenHands issue resolution agent augmented with oracle locations (\S\ref{sec:downstream_swebench}).
\label{fig:user_prompt_oracle_augment2}
}
\end{figure*}

\begin{figure*}[!ht]
    \centering
\begin{tcolorbox}[colback=green!6!white, colframe=green!75!black, title=Default OpenHands System Prompt used for Issue Resolution (continued on next page)]
\begin{lstlisting}
You are OpenHands agent, a helpful AI assistant that can interact with a computer to solve tasks.

<ROLE>
* Your primary role is to assist users by executing commands, modifying code, and solving technical problems effectively. You should be thorough, methodical, and prioritize quality over speed.
* If the user asks a question, like \"why is X happening\", don't try to fix the problem. Just give an answer to the question.
</ROLE>

<MEMORY>
* Use `AGENTS.md` under the repository root as your persistent memory for repository-specific knowledge and context.
* Add important insights, patterns, and learnings to this file to improve future task performance.
* This repository skill is automatically loaded for every conversation and helps maintain context across sessions.
* For more information about skills, see: https://docs.openhands.dev/overview/skills
</MEMORY>

<EFFICIENCY>
* Each action you take is somewhat expensive. Wherever possible, combine multiple actions into a single action, e.g. combine multiple bash commands into one, using sed and grep to edit/view multiple files at once.
* When exploring the codebase, use efficient tools like find, grep, and git commands with appropriate filters to minimize unnecessary operations.
</EFFICIENCY>

<FILE_SYSTEM_GUIDELINES>
* When a user provides a file path, do NOT assume it's relative to the current working directory. First explore the file system to locate the file before working on it.
* If asked to edit a file, edit the file directly, rather than creating a new file with a different filename.
* For global search-and-replace operations, consider using `sed` instead of opening file editors multiple times.
* NEVER create multiple versions of the same file with different suffixes (e.g., file_test.py, file_fix.py, file_simple.py). Instead:
  - Always modify the original file directly when making changes
  - If you need to create a temporary file for testing, delete it once you've confirmed your solution works
  - If you decide a file you created is no longer useful, delete it instead of creating a new version
* Do NOT include documentation files explaining your changes in version control unless the user explicitly requests it
* When reproducing bugs or implementing fixes, use a single file rather than creating multiple files with different versions
</FILE_SYSTEM_GUIDELINES>
\end{lstlisting}
\end{tcolorbox}
\caption{System prompt of the OpenHands issue resolution agent used for experiments in \S\ref{sec:downstream_swebench}.}
\label{fig:swe_bench_system_prompt1}
\end{figure*}

\begin{figure*}[!h]
    \centering
\begin{tcolorbox}[colback=green!6!white, colframe=green!75!black, title=Default OpenHands System Prompt used for Issue Resolution (continued on next page)]
\begin{lstlisting}
<CODE_QUALITY>
* Write clean, efficient code with minimal comments. Avoid redundancy in comments: Do not repeat information that can be easily inferred from the code itself.
* When implementing solutions, focus on making the minimal changes needed to solve the problem.
* Before implementing any changes, first thoroughly understand the codebase through exploration.
* If you are adding a lot of code to a function or file, consider splitting the function or file into smaller pieces when appropriate.
* Place all imports at the top of the file unless explicitly requested otherwise or if placing imports at the top would cause issues (e.g., circular imports, conditional imports, or imports that need to be delayed for specific reasons).
</CODE_QUALITY>

<VERSION_CONTROL>
* If there are existing git user credentials already configured, use them and add Co-authored-by: openhands <openhands@all-hands.dev> to any commits messages you make. if a git config doesn't exist use \"openhands\" as the user.name and \"openhands@all-hands.dev\" as the user.email by default, unless explicitly instructed otherwise.
* Exercise caution with git operations. Do NOT make potentially dangerous changes (e.g., pushing to main, deleting repositories) unless explicitly asked to do so.
* When committing changes, use `git status` to see all modified files, and stage all files necessary for the commit. Use `git commit -a` whenever possible.
* Do NOT commit files that typically shouldn't go into version control (e.g., node_modules/, .env files, build directories, cache files, large binaries) unless explicitly instructed by the user.
* If unsure about committing certain files, check for the presence of .gitignore files or ask the user for clarification.
* When running git commands that may produce paged output (e.g., `git diff`, `git log`, `git show`), use `git --no-pager <command>` or set `GIT_PAGER=cat` to prevent the command from getting stuck waiting for interactive input.
</VERSION_CONTROL>

<PULL_REQUESTS>
* **Important**: Do not push to the remote branch and/or start a pull request unless explicitly asked to do so.
* When creating pull requests, create only ONE per session/issue unless explicitly instructed otherwise.
* When working with an existing PR, update it with new commits rather than creating additional PRs for the same issue.
* When updating a PR, preserve the original PR title and purpose, updating description only when necessary.
</PULL_REQUESTS>

<PROBLEM_SOLVING_WORKFLOW>
1. EXPLORATION: Thoroughly explore relevant files and understand the context before proposing solutions
2. ANALYSIS: Consider multiple approaches and select the most promising one
3. TESTING:
   * For bug fixes: Create tests to verify issues before implementing fixes
   * For new features: Consider test-driven development when appropriate
   * Do NOT write tests for documentation changes, README updates, configuration files, or other non-functionality changes
   * Do not use mocks in tests unless strictly necessary and justify their use when they are used. You must always test real code paths in tests, NOT mocks.
   * If the repository lacks testing infrastructure and implementing tests would require extensive setup, consult with the user before investing time in building testing infrastructure
\end{lstlisting}
\end{tcolorbox}
\caption{System prompt of the OpenHands issue resolution agent used for experiments in \S\ref{sec:downstream_swebench}.}
\label{fig:swe_bench_system_prompt2}
\end{figure*}

\begin{figure*}[!h]
    \centering
\begin{tcolorbox}[colback=green!6!white, colframe=green!75!black, title=Default OpenHands System Prompt used for Issue Resolution (continued on next page)]
\begin{lstlisting}
   * If the environment is not set up to run tests, consult with the user first before investing time to install all dependencies
4. IMPLEMENTATION:
   * Make focused, minimal changes to address the problem
   * Always modify existing files directly rather than creating new versions with different suffixes
   * If you create temporary files for testing, delete them after confirming your solution works
5. VERIFICATION: If the environment is set up to run tests, test your implementation thoroughly, including edge cases. If the environment is not set up to run tests, consult with the user first before investing time to run tests.
</PROBLEM_SOLVING_WORKFLOW>

<SELF_DOCUMENTATION>
When the user directly asks about any of the following:
- OpenHands capabilities (e.g., \"can OpenHands do...\", \"does OpenHands have...\")
- what you're able to do in second person (e.g., \"are you able...\", \"can you...\")
- how to use a specific OpenHands feature or product
- how to use the OpenHands SDK, CLI, GUI, or other OpenHands products

Get accurate information from the official OpenHands documentation at <https://docs.openhands.dev/>. The documentation includes:

**OpenHands SDK** (`/sdk/*`): Python library for building AI agents; Getting Started, Architecture, Guides (agent, llm, conversation, tools), API Reference
**OpenHands CLI** (`/openhands/usage/run-openhands/cli-mode`): Command-line interface
**OpenHands GUI** (`/openhands/usage/run-openhands/local-setup`): Local GUI and REST API
**OpenHands Cloud** (`/openhands/usage/run-openhands/cloud`): Hosted solution with integrations
**OpenHands Enterprise**: Self-hosted deployment with extended support

Always provide links to the relevant documentation pages for users who want to learn more.
</SELF_DOCUMENTATION>

<SECURITY>
# Security Policy

## OK to do without Explicit User Consent

- Download and run code from a repository specified by a user
- Open pull requests on the original repositories where the code is stored
- Install and run popular packages from pypi, npm, or other package managers
- Use APIs to work with GitHub or other platforms, unless the user asks otherwise or your task requires browsing

## Do only with Explicit User Consent

- Upload code to anywhere other than the location where it was obtained from
- Upload API keys or tokens anywhere, except when using them to authenticate with the appropriate service
\end{lstlisting}
\end{tcolorbox}
\caption{System prompt of the OpenHands issue resolution agent used for experiments in \S\ref{sec:downstream_swebench}.}
\label{fig:swe_bench_system_prompt3}
\end{figure*}

\begin{figure*}[!h]
    \centering
\begin{tcolorbox}[colback=green!6!white, colframe=green!75!black, title=Default OpenHands System Prompt used for Issue Resolution (continued on next page)]
\begin{lstlisting}
## Never Do

- Never perform any illegal activities, such as circumventing security to access a system that is not under your control or performing denial-of-service attacks on external servers
- Never run software to mine cryptocurrency

## General Security Guidelines

- Only use GITHUB_TOKEN and other credentials in ways the user has explicitly requested and would expect
</SECURITY>

<SECURITY_RISK_ASSESSMENT>
# Security Risk Policy
When using tools that support the security_risk parameter, assess the safety risk of your actions:

- **LOW**: Safe, read-only actions.
  - Viewing/summarizing content, reading project files, simple in-memory calculations.
- **MEDIUM**: Project-scoped edits or execution.
  - Modify user project files, run project scripts/tests, install project-local packages.
- **HIGH**: System-level or untrusted operations.
  - Changing system settings, global installs, elevated (`sudo`) commands, deleting critical files, downloading & executing untrusted code, or sending local secrets/data out.

**Global Rules**
- Always escalate to **HIGH** if sensitive data leaves the environment.
</SECURITY_RISK_ASSESSMENT>

<EXTERNAL_SERVICES>
* When interacting with external services like GitHub, GitLab, or Bitbucket, use their respective APIs instead of browser-based interactions whenever possible.
* Only resort to browser-based interactions with these services if specifically requested by the user or if the required operation cannot be performed via API.
</EXTERNAL_SERVICES>

<ENVIRONMENT_SETUP>
* When user asks you to run an application, don't stop if the application is not installed. Instead, please install the application and run the command again.
* If you encounter missing dependencies:
  1. First, look around in the repository for existing dependency files (requirements.txt, pyproject.toml, package.json, Gemfile, etc.)
  2. If dependency files exist, use them to install all dependencies at once (e.g., `pip install -r requirements.txt`, `npm install`, etc.)
  3. Only install individual packages directly if no dependency files are found or if only specific packages are needed
* Similarly, if you encounter missing dependencies for essential tools requested by the user, install them when possible.
</ENVIRONMENT_SETUP>
\end{lstlisting}
\end{tcolorbox}
\caption{System prompt of the OpenHands issue resolution agent used for experiments in \S\ref{sec:downstream_swebench}.}
\label{fig:swe_bench_system_prompt4}
\end{figure*}
\begin{figure*}[!htpb]
    \centering
\begin{tcolorbox}[colback=green!6!white, colframe=green!75!black, title=Default OpenHands System Prompt used for Issue Resolution]
\begin{lstlisting}
<TROUBLESHOOTING>
* If you've made repeated attempts to solve a problem but tests still fail or the user reports it's still broken:
  1. Step back and reflect on 5-7 different possible sources of the problem
  2. Assess the likelihood of each possible cause
  3. Methodically address the most likely causes, starting with the highest probability
  4. Explain your reasoning process in your response to the user
* When you run into any major issue while executing a plan from the user, please don't try to directly work around it. Instead, propose a new plan and confirm with the user before proceeding.
</TROUBLESHOOTING>

<PROCESS_MANAGEMENT>
* When terminating processes:
  - Do NOT use general keywords with commands like `pkill -f server` or `pkill -f python` as this might accidentally kill other important servers or processes
  - Always use specific keywords that uniquely identify the target process
  - Prefer using `ps aux` to find the exact process ID (PID) first, then kill that specific PID
  - When possible, use more targeted approaches like finding the PID from a pidfile or using application-specific shutdown commands
</PROCESS_MANAGEMENT>
\end{lstlisting}
\end{tcolorbox}
\caption{System prompt of the OpenHands issue resolution agent used for experiments in \S\ref{sec:downstream_swebench}.}
\label{fig:swe_bench_system_prompt5}\end{figure*}

%% file: Figures/codescout_14b_reward_fig.tex
\begin{figure}[htbp]
     \centering
     \begin{subfigure}[b]{0.48\textwidth}
         \centering
         \includegraphics[width=\textwidth]{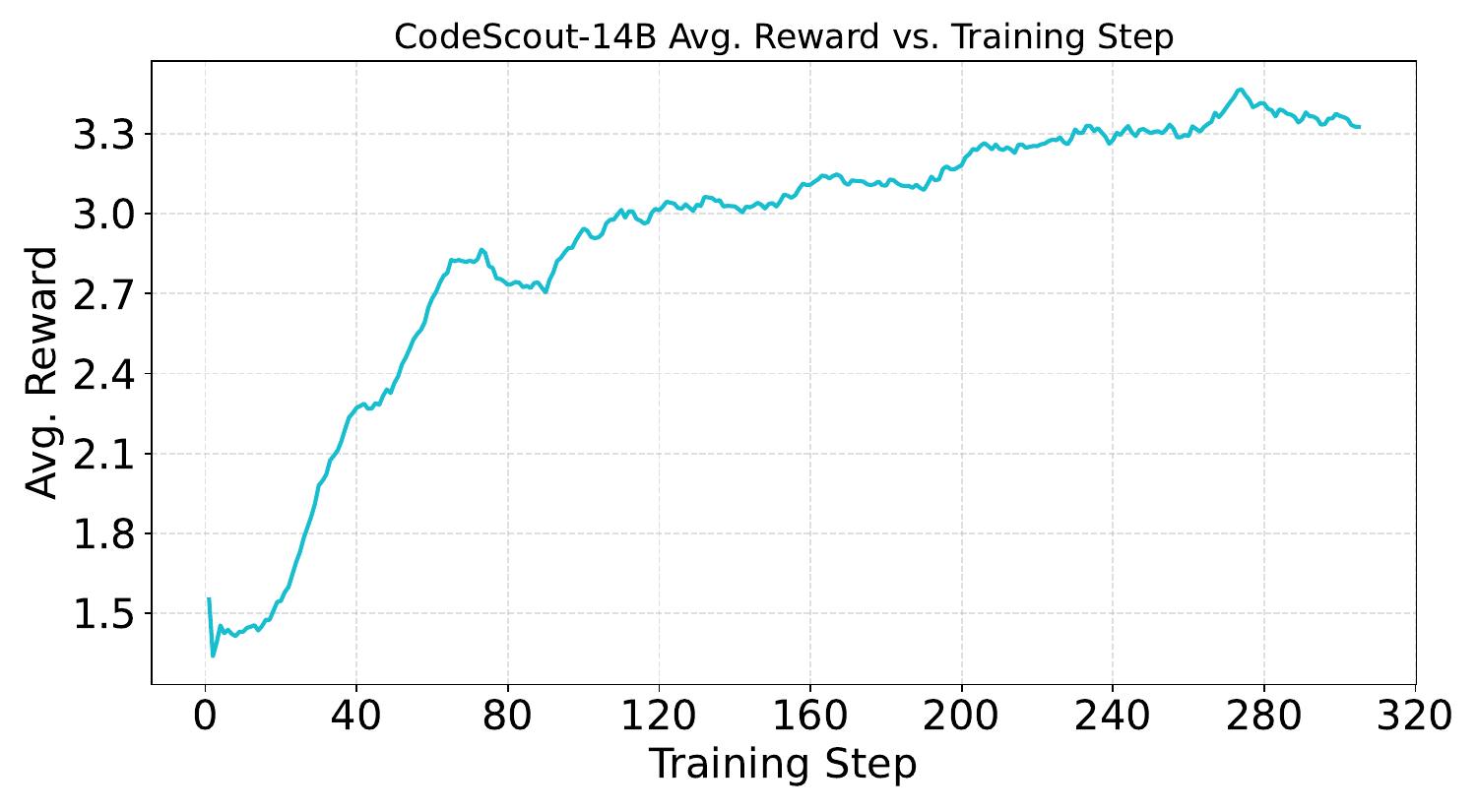}
         \caption{Aggregate reward vs. Training Step}
         \label{fig:codescout_14b_avg}
     \end{subfigure}
     \begin{subfigure}[b]{0.48\textwidth}
         \centering
         \includegraphics[width=\textwidth]{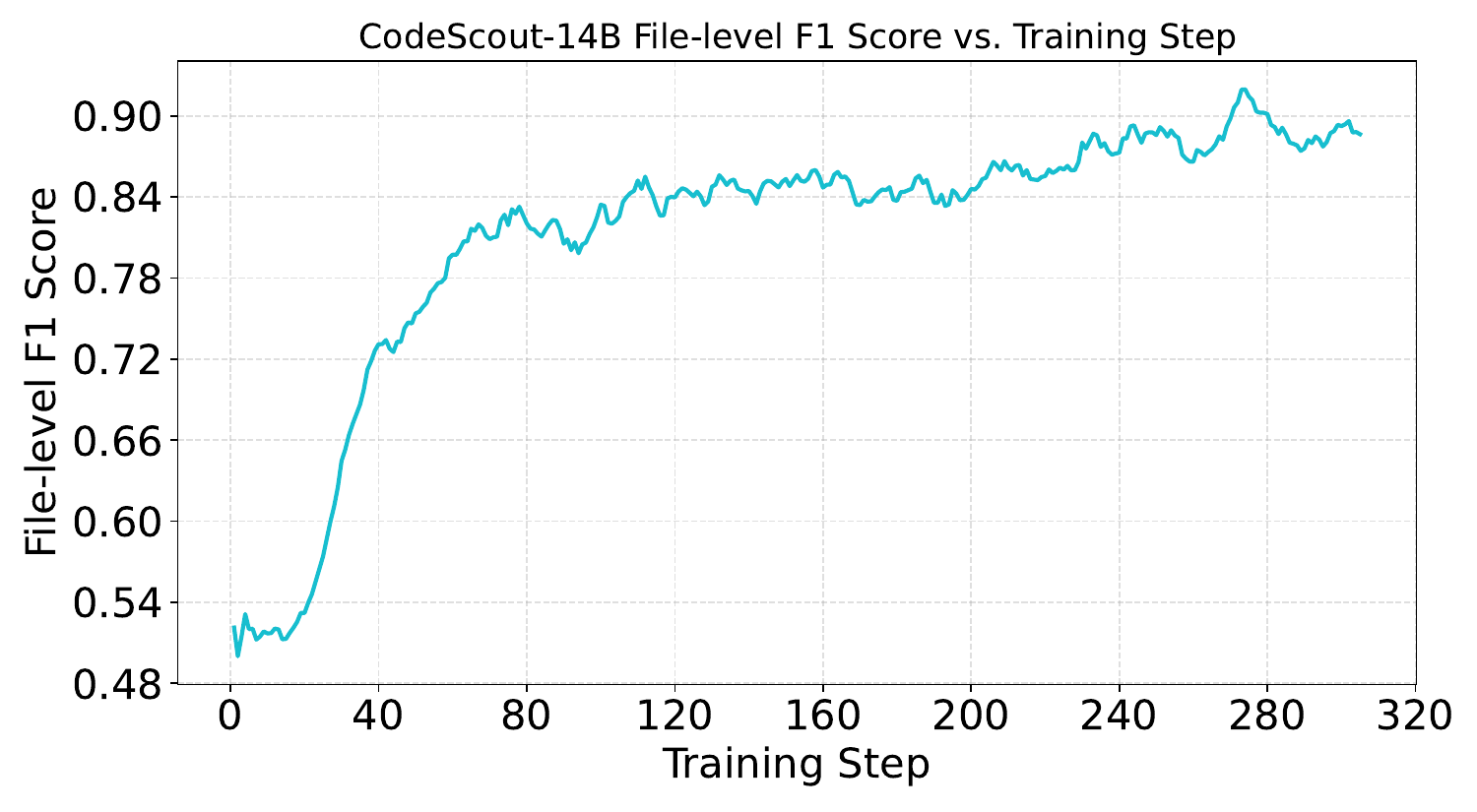}
         \caption{File-level F1 score vs. Training Step}
         \label{fig:codescout_14b_file}
     \end{subfigure}

     \vspace{5pt} %

     \begin{subfigure}[b]{0.48\textwidth}
         \centering
         \includegraphics[width=\textwidth]{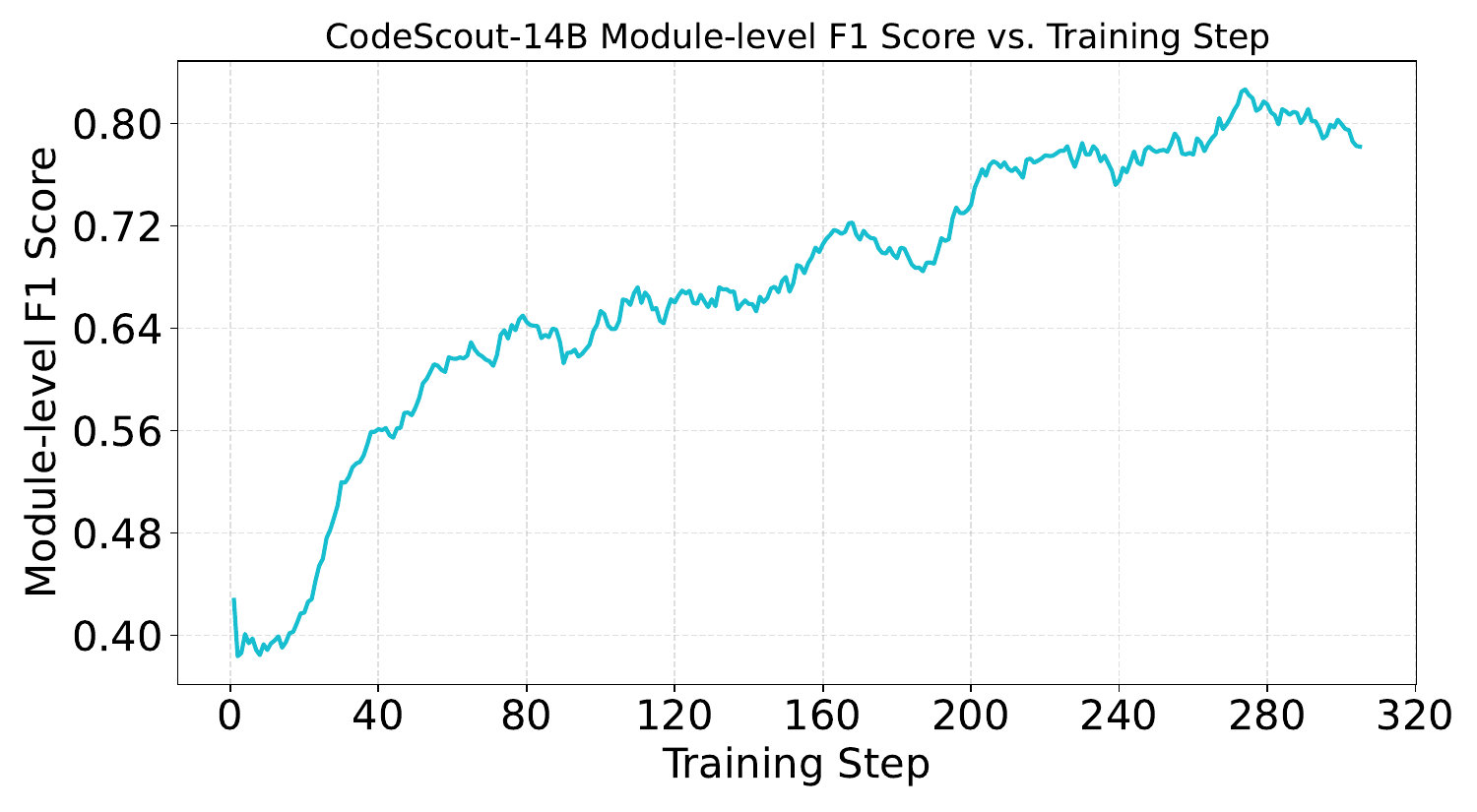}
         \caption{Module-level F1 score vs. Training Step}
         \label{fig:codescout_14b_module}
     \end{subfigure}
     \hfill
     \begin{subfigure}[b]{0.48\textwidth}
         \centering
         \includegraphics[width=\textwidth]{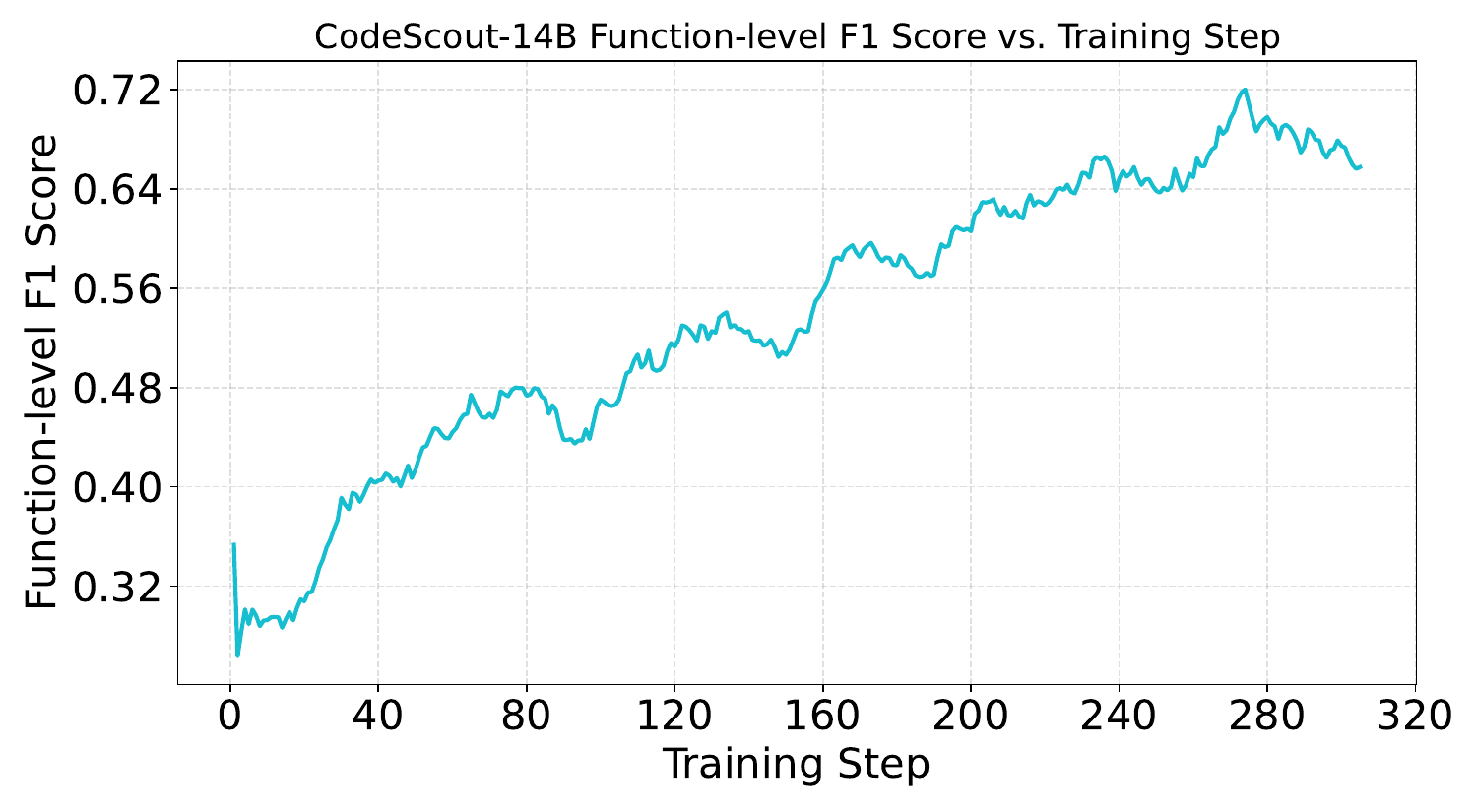}
         \caption{Function-level F1 score vs. Training Step}
         \label{fig:codescout_14b_function}
     \end{subfigure}
     
     \caption{Reward curves for RL training of \methodname-14B.}
     \label{fig:codescout_14b_reward}
\end{figure}

%% file: Figures/codescout_4b_reward_fig.tex
\begin{figure}[htbp]
     \centering
     \begin{subfigure}[b]{0.48\textwidth}
         \centering
         \includegraphics[width=\textwidth]{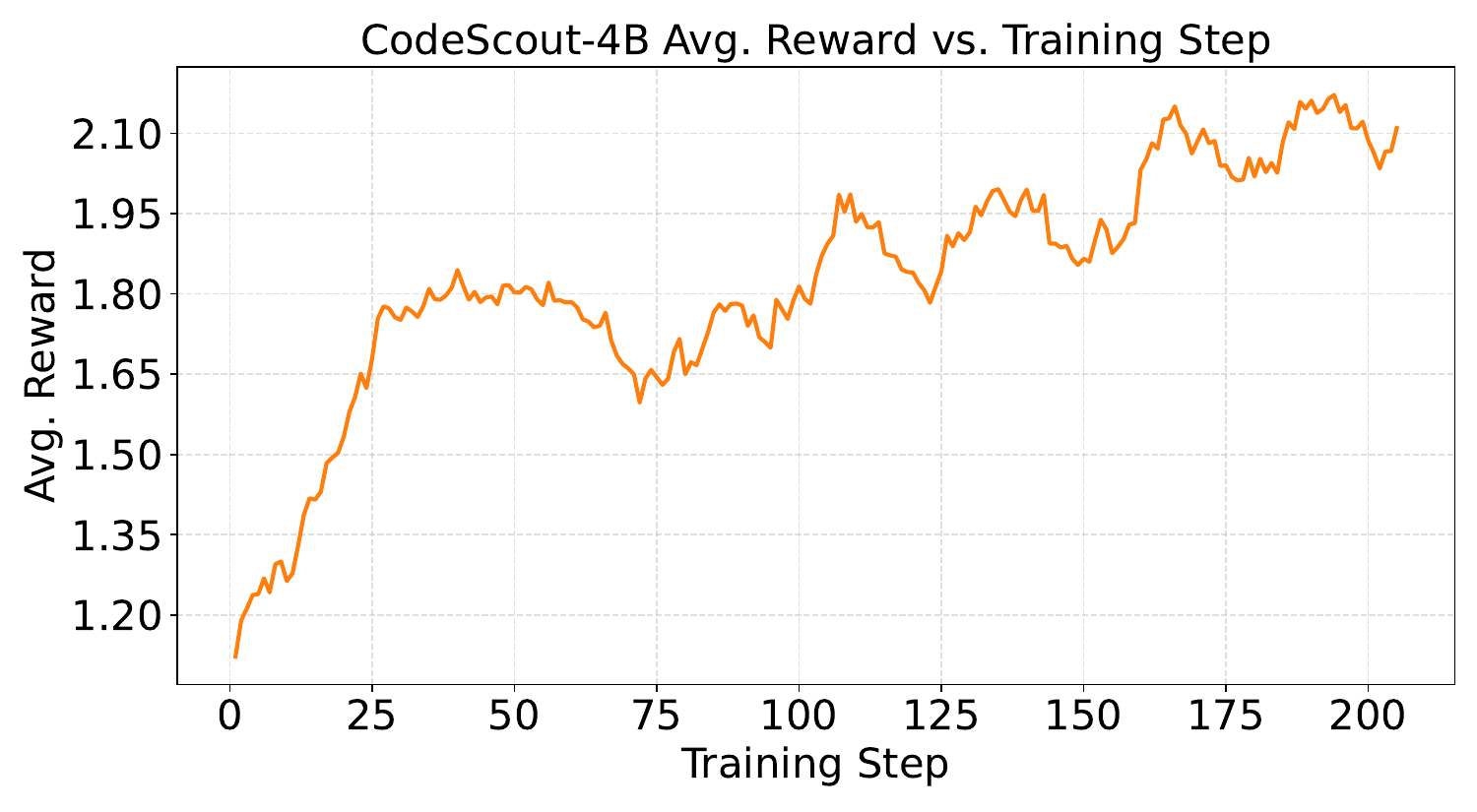}
         \caption{Aggregate reward vs. Training Step}
         \label{fig:codescout_4b_avg}
     \end{subfigure}
     \hfill
     \begin{subfigure}[b]{0.48\textwidth}
         \centering
         \includegraphics[width=\textwidth]{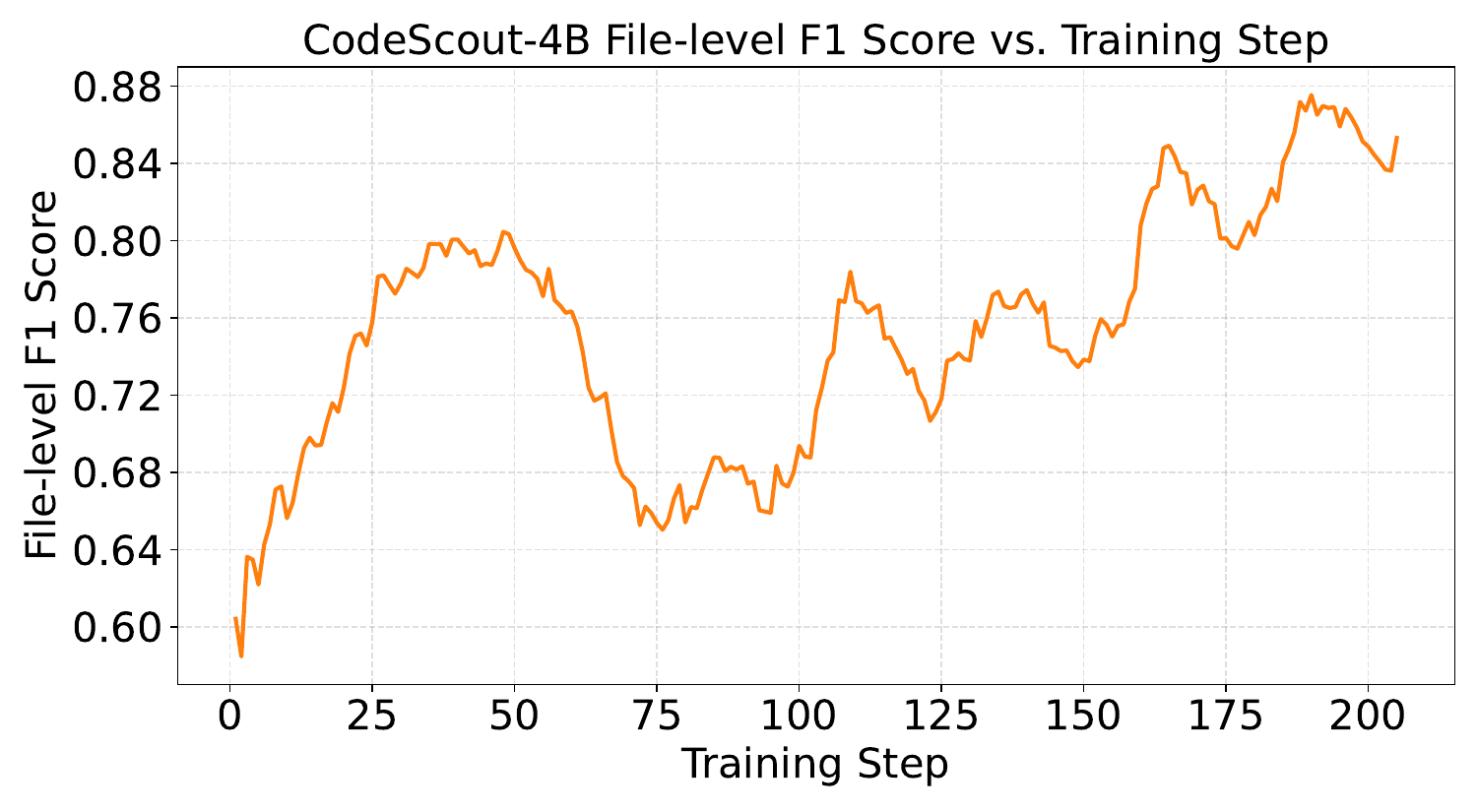}
         \caption{File-level F1 score vs. Training Step}
         \label{fig:image2}
     \end{subfigure}

     \vspace{5pt} %

     \begin{subfigure}[b]{0.48\textwidth}
         \centering
         \includegraphics[width=\textwidth]{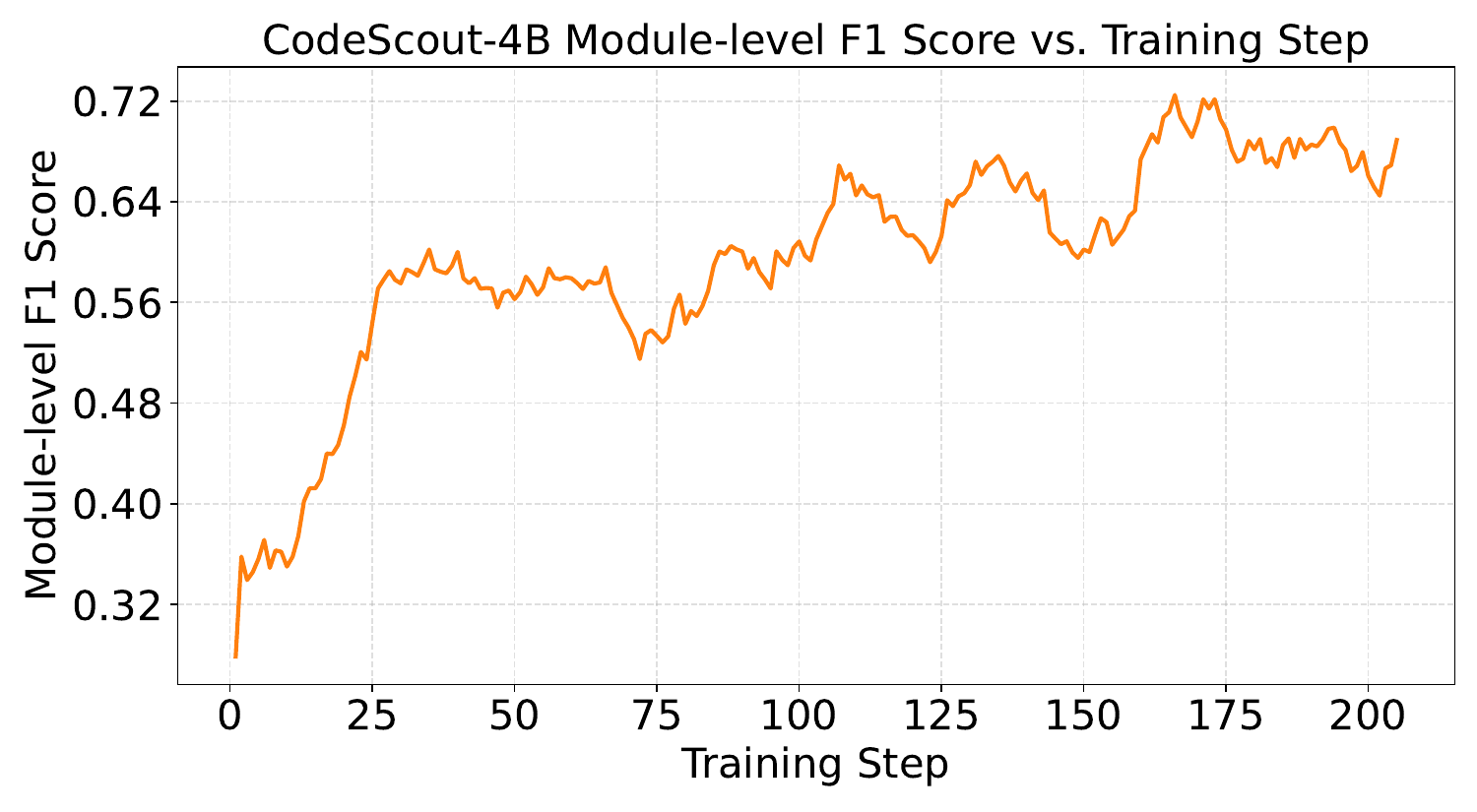}
         \caption{Module-level F1 score vs. Training Step}
         \label{fig:image3}
     \end{subfigure}
     \hfill
     \begin{subfigure}[b]{0.48\textwidth}
         \centering
         \includegraphics[width=\textwidth]{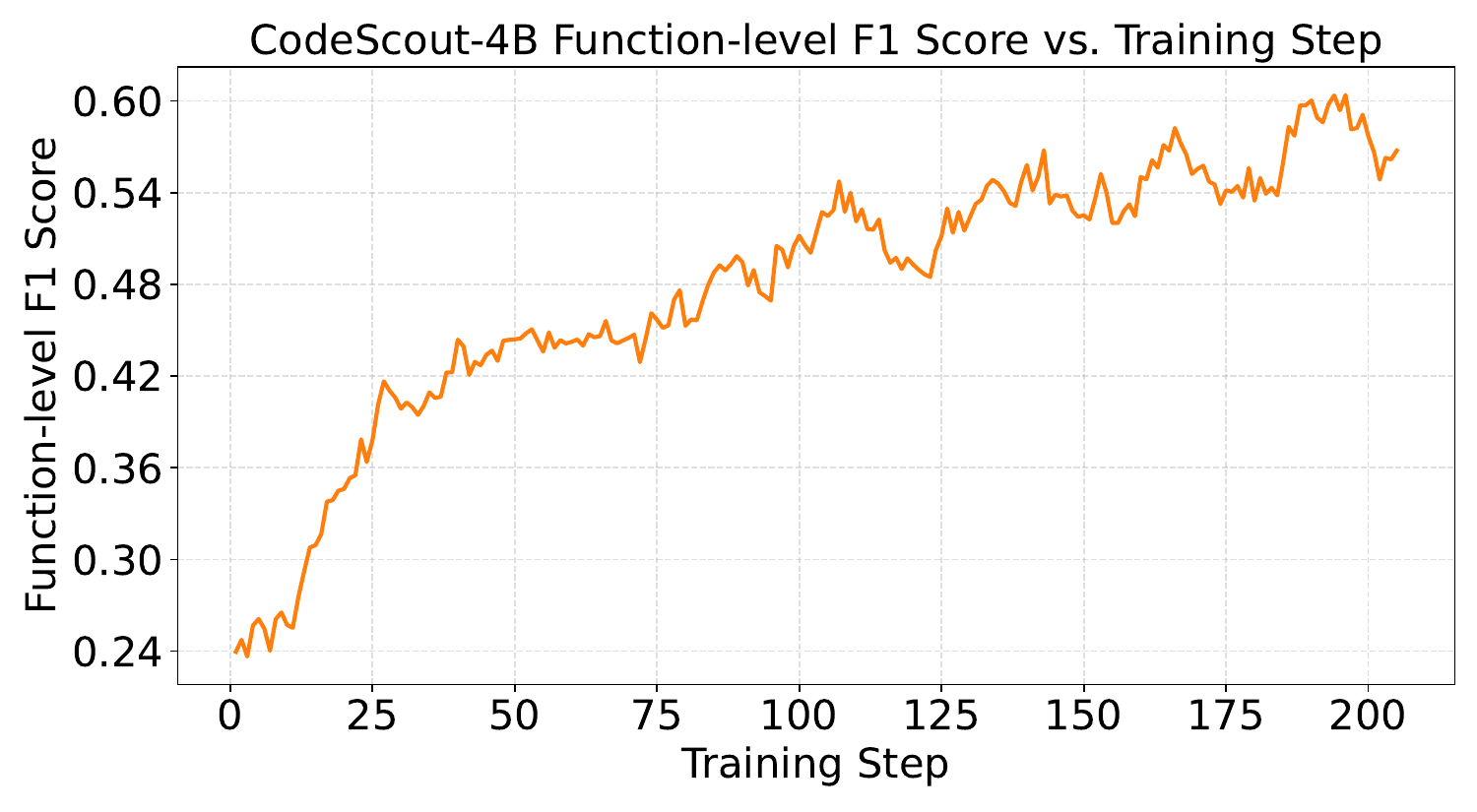}
         \caption{Function-level F1 score vs. Training Step}
         \label{fig:image4}
     \end{subfigure}
     
     \caption{Reward curves for RL training of \methodname-4B.}
     \label{fig:codescout_4b_reward}
\end{figure}

%% file: Figures/codescout_1.7b_reward_fig.tex
\begin{figure}[htbp]
     \centering
     \begin{subfigure}[b]{0.48\textwidth}
         \centering
         \includegraphics[width=\textwidth]{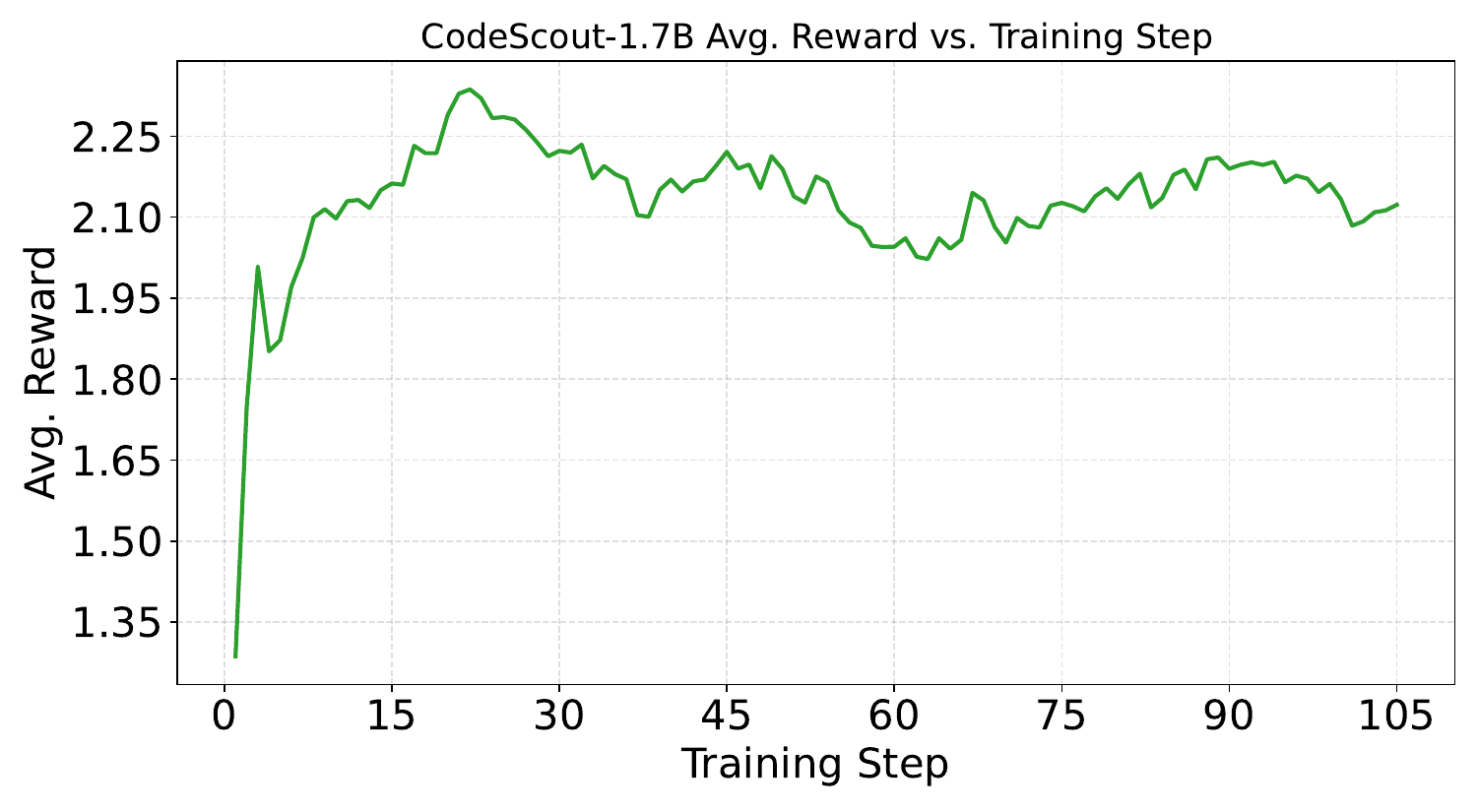}
         \caption{Aggregate reward vs. Training Step}
         \label{fig:codescout_1.7b_avg}
     \end{subfigure}
     \begin{subfigure}[b]{0.48\textwidth}
         \centering
         \includegraphics[width=\textwidth]{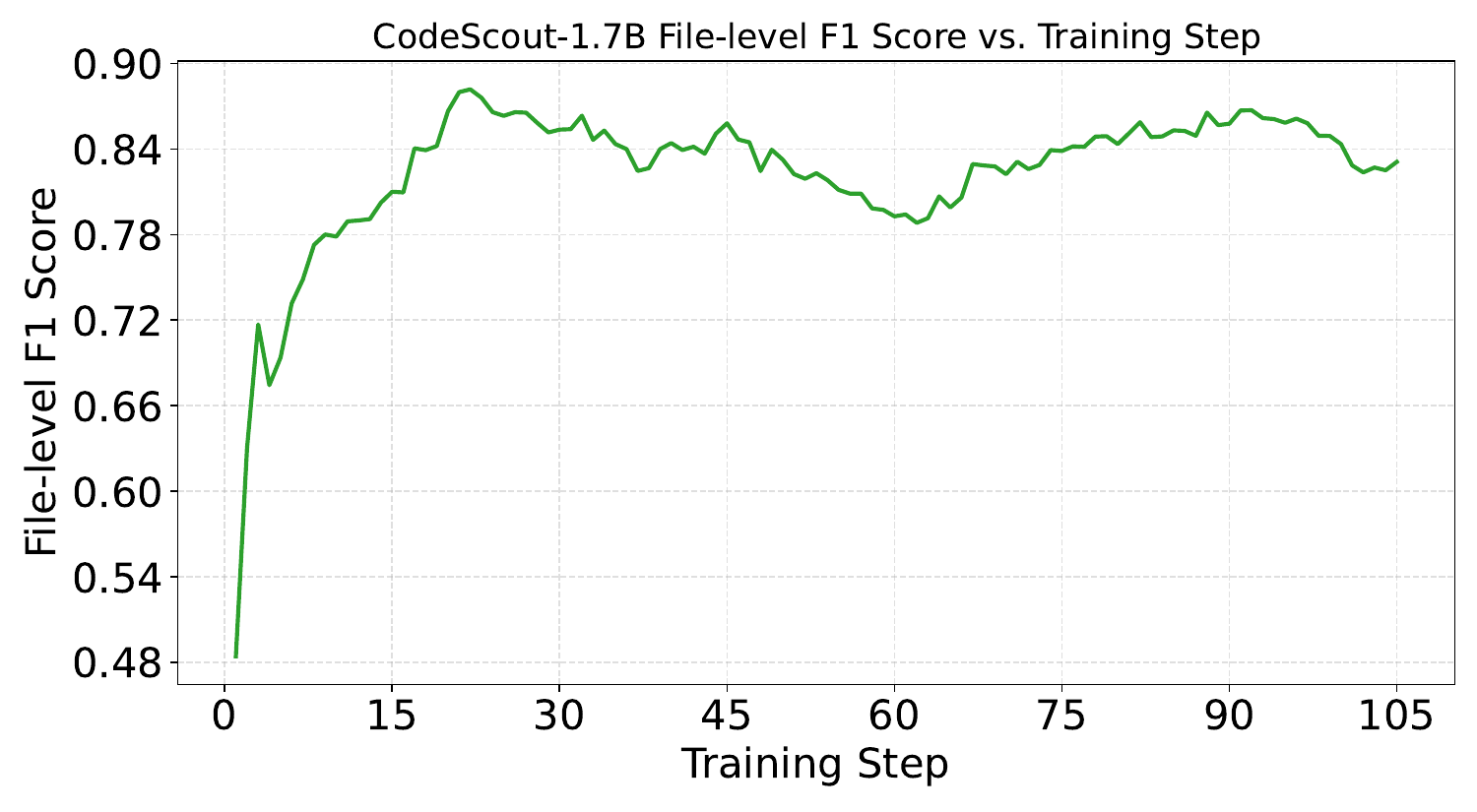}
         \caption{File-level F1 score vs. Training Step}
         \label{fig:codescout_1.7b_file}
     \end{subfigure}

     \vspace{5pt} %

     \begin{subfigure}[b]{0.48\textwidth}
         \centering
         \includegraphics[width=\textwidth]{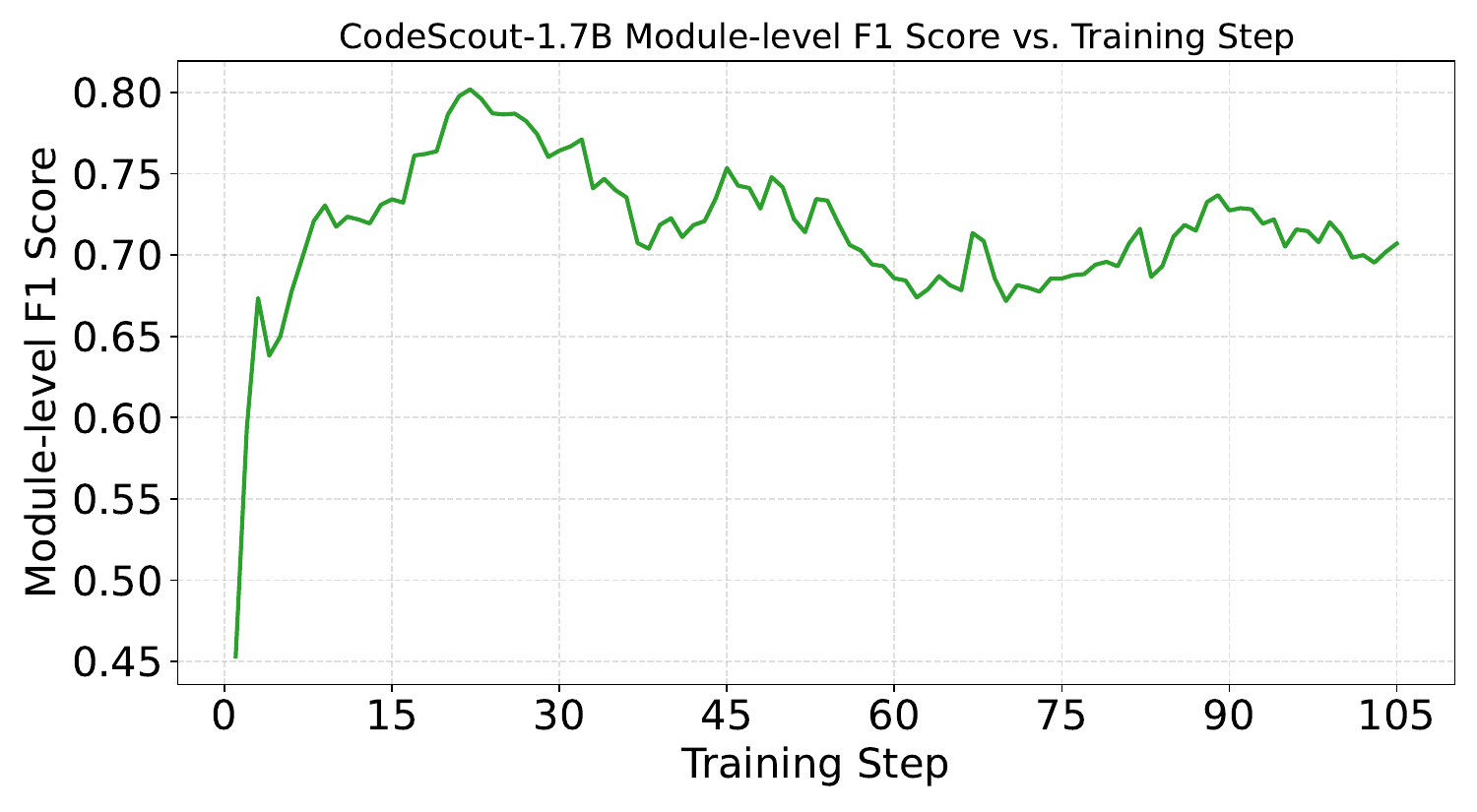}
         \caption{Module-level F1 score vs. Training Step}
         \label{fig:codescout_1.7b_module}
     \end{subfigure}
     \hfill
     \begin{subfigure}[b]{0.48\textwidth}
         \centering
         \includegraphics[width=\textwidth]{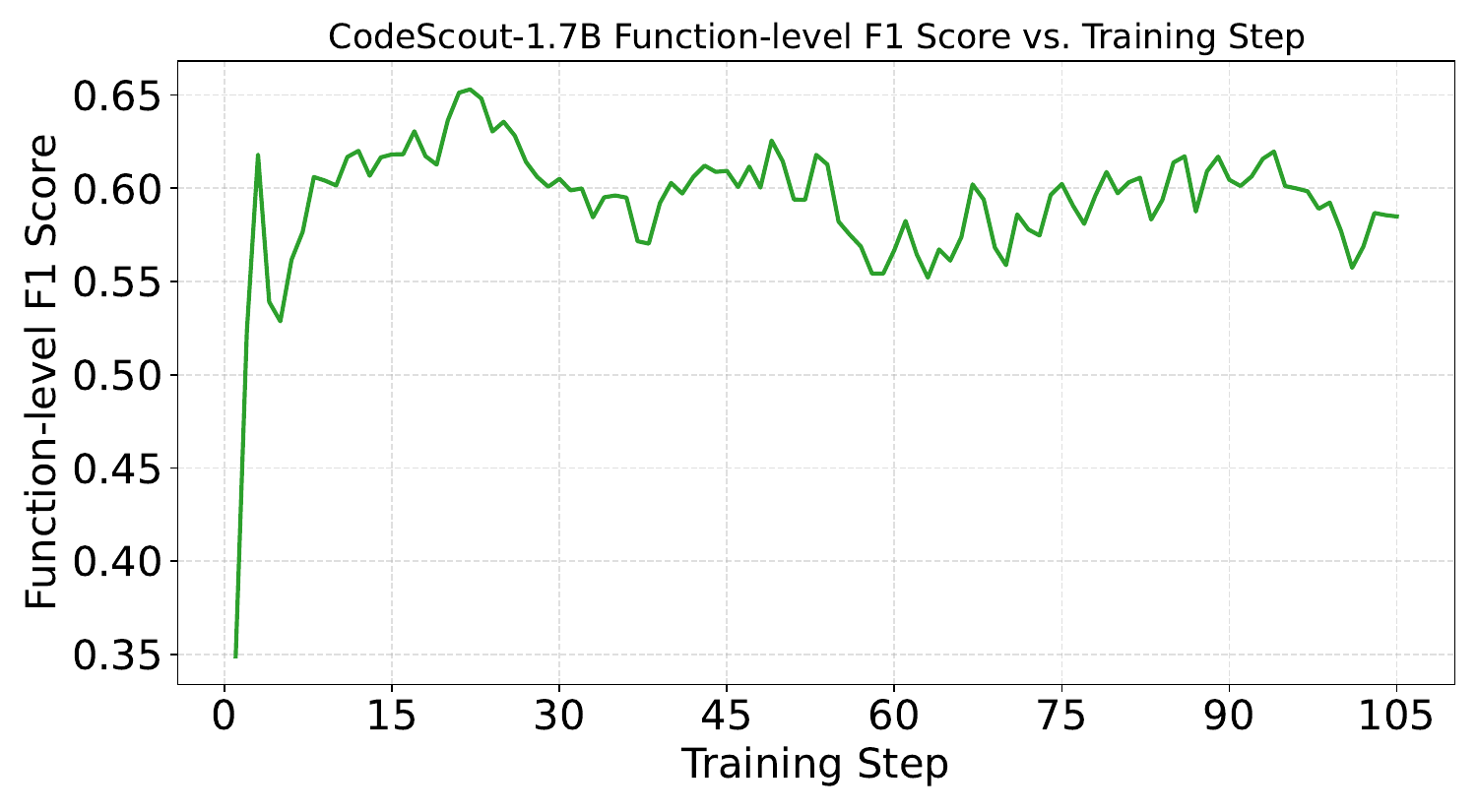}
         \caption{Function-level F1 score vs. Training Step}
         \label{fig:codescout_1.7b_function}
     \end{subfigure}
     
     \caption{Reward curves for RL training of \methodname-1.7B.}
     \label{fig:codescout_1.7b_reward}
\end{figure}

%% file: Tables/baselines_tool_descriptions.tex
\begin{table}[!t]
\centering
\caption{Comparison of tools used by prior code localization agents and \methodname. Existing approaches rely on specialized task-specific tools, while \methodname achieves competitive or superior performance using only a bash terminal typical of coding agents.}
\label{tab:granular_tool_behaviours}
\renewcommand{\arraystretch}{1.2}
\resizebox{\columnwidth}{!}{%
\begin{tabular}{lll}
\toprule
\textbf{Method} & \textbf{Tools} & \textbf{Tool Behaviour} \\
\midrule
\multirow{3}{*}{LocAgent~\citep{chen-etal-2025-locagent}} & \texttt{SearchEntity} & Performs keyword-based retrieval to identify relevant code entities. \\
 & \texttt{TraverseGraph} & Traverse/parse codebase dependencies by navigating hops in the code graph. \\
 & \texttt{RetrieveEntity} & Accesses the complete implementation for a specific entity ID. \\
 \midrule
\multirow{3}{*}{CoSIL~\citep{liu2025cosil}} & \texttt{search\_class\_node} & Retrieves the full code snippet of a class given its file path. \\
 & \texttt{search\_class\_function\_node} & Extracts implementation details for a specific member function in a class. \\
 & \texttt{search\_file\_function\_node} & Fetches the source code for standalone/static functions within a file. \\
\midrule
\multirow{5}{*}{OrcaLoca~\citep{yu2025orcalocallmagentframework}} & \texttt{search\_file\_contents} & Returns file contents or a structural skeleton for long files ($>$200 lines). \\
 & \texttt{search\_class} & Locates and returns class definitions or their structural skeletons. \\
 & \texttt{search\_method\_in\_class} & Extracts specific method implementations from within a designated class. \\
 & \texttt{search\_callable} & Identifies and returns code snippets for callable objects (functions/methods). \\
 & \texttt{search\_source\_code} & Performs a general search to match source code strings to snippets. \\
\midrule
\multirow{5}{*}{RepoSearcher~\citep{ma2025toolintegratedreinforcementlearningrepo}} & \texttt{GetRepoStructure} & Provides an overview of the repository’s file and directory hierarchy. \\
 & \texttt{GetImportOfFile} & Identifies and lists all imports for a given file. \\
 & \texttt{SearchClass} & Fetches the raw code content of a specific class definition. \\
 & \texttt{SearchFunction} & Fetches the raw code content of a standalone function definition. \\
 & \texttt{SearchClassMethod} & Specifically targets and retrieves implementations of class-specific methods. \\
\midrule
RepoNavigator~\citep{zhang2025one} & \texttt{jump} & Navigates directly to a symbol's definition using a language server. \\
\midrule
\rowcolor{lightgolden}\textbf{\methodname \textit{(Ours)}} & \texttt{terminal} & Executes standard Unix commands in a persistent, stateful \texttt{tmux} session \\
\bottomrule
\end{tabular}%
}
\end{table}

%% file: Sections/7_related_work.tex
\subsection{Approaches for code localization}
Several prior methods have developed methods targeting code localization for downstream issue resolution in repositories. \citet{jimenez2024swebenchlanguagemodelsresolve} identify relevant source code files using BM25 retrieval~\citep{article} treating the issue description as the query and the Python source code files as documents. SWE-Fixer~\citep{xie2025swefixertrainingopensourcellms} use a two-step coarse-to-fine localization approach where they first retrieve relevant source code files using BM25 retrieval and then prompt a fine-tuned 7B model with the skeleton of the retrieved files to predict the names of relevant files for issue resolution. On the other hand, Agentless~\citep{xia2024agentless} uses a complex, multi-step localization pipeline wherein it first retrieves suspicious code files from the repository using both embedding-based retrieval and by prompting an LLM with the high-level repository structure, followed by prompting an LLM to predict relevant functions and classes given the skeletons of the suspicious files. 

Many recent approaches have increasingly shifted towards developing models and systems for code localization using agentic scaffolds. Table~\ref{tab:granular_tool_behaviours} provides an overview of tools used by several prior agentic approaches and \methodname for code localization. LocAgent~\citep{chen-etal-2025-locagent} and RepoGraph~\citep{ouyang2025repographenhancingaisoftware} utilize a graph-based indexing approach wherein all dependencies in the code repository (for e.g. import, invoke, inherit, etc.) are captured in a code graph and also develop specialized tools allowing the agent to search and traverse the graph. Furthermore, \citet{chen-etal-2025-locagent} train a 7B model for their scaffold using rejection sampling fine-tuning on successful trajectories sampled from Claude-3.5-Sonnet. On the other hand, CoSIL~\citep{liu2025cosil} first localizes relevant files using module-call graphs and then identifies relevant functions using function-call graphs. Note that these call graphs are constructed on-the-fly during inference as opposed to pre-indexing the code graph done by ~\citet{chen-etal-2025-locagent}. OrcaLoca~\citep{yu2025orcalocallmagentframework} performs code localization through efficient exploration of the code graph using priority-based action scheduling, action decomposition with relevance scoring, and distance-aware context pruning. Note that many of these scaffolds are not suitable for reinforcement learning as they often require expensive repository pre-processing (for e.g. creating a code graph) increasing the computational overhead of performing rollouts during reinforcement learning.

RepoSearcher~\citep{ma2025toolintegratedreinforcementlearningrepo} develop a light-weight scaffold with tools specialized for localization that allow searching for classes and functions in files, obtaining imports of a given file, etc. They use a two-stage approach to train open-source LLMs for their scaffold: (1) rejection sampling fine-tuning from a closed-source LLM (Claude3.7-Sonnet) to warm-up their model, (2) reinforcement learning on the fine-tuned checkpoint to further enhance performance. RepoNavigator~\citep{zhang2025one} use an even simpler scaffold comprising with just one tool - \texttt{jump} - which allows the agent to retrieve definitions of Python symbols in files. Although~\citet{zhang2025one} do not rely on rejection sampling fine-tuning from closed-source LLMs and directly train models using reinforcement learning, they still rely on selecting ``easy" training instances by discarding all training instances that were not successfully solved by the base Qwen2.5-7B model equipped with their scaffold atleast once among 16 sampled trajectories. 

\subsection{Training methods for software engineering agents}
Several prior approaches have trained LLM-based software engineering agents for various downstream tasks like code localization and issue resolution. Many methods rely on performing rejection sampling fine-tuning~\citep{yuan2023scalingrelationshiplearningmathematical}: sample trajectories (for training instances) from a stronger (often closed-source) model and train a smaller model on trajectories which successfully solve the task. Prior approaches that use this approach include LocAgent~\citep{chen-etal-2025-locagent} and RepoSearcher~\citep{ma2025toolintegratedreinforcementlearningrepo} for code localization, and SWE-Gym~\citep{pan2025trainingsoftwareengineeringagents}, SWE-Smith~\citep{yang2025swesmithscalingdatasoftware}, and R2E-Gym~\citep{jain2025r2egymproceduralenvironmentshybrid} for issue resolution. Recently, various attempts have been made to train models directly with reinforcement learning without relying on proprietary, closed-source LLMs for rejection sampling fine-tuning. While RepoNavigator~\citep{zhang2025one}, SWE-Grep~\citep{cognition2025swegrep}, and SID-1~\citep{sid2025preview} train LLM agents with RL for code localization, DeepSWE~\citep{luo2025deepswe} and SkyRL-v0~\citep{cao2025skyrl} leverage RL to train LLM agents for issue resolution.

%% file: Tables/results_train_ablation_detailed.tex
\begin{table}[!t]
\centering
\caption{Ablation study on RL algorithm variants for \texttt{Qwen3-4B-Instruct-2507}, evaluated on SWE-Bench Pro.}
\label{tab:results_ablation}
\renewcommand{\arraystretch}{1.3}
\setlength{\tabcolsep}{3pt}
\resizebox{\textwidth}{!}{%
\begin{tabular}{llccc}
\toprule
\textbf{Training Algorithm} & \textbf{Key Configuration} & \textbf{File F1} & \textbf{Module F1} & \textbf{Function F1} \\
\midrule
\multicolumn{5}{l}{\textit{Recipe-level comparison}} \\
\midrule
GSPO
  & Seq-level ratio, tight clip ($\epsilon$=3e-4/4e-4), seq-mean reduction, no std norm
  & \textbf{54.83\%} & 31.43\% & 23.29\% \\
SAPO
  & Soft gating, seq-mean reduction, std norm
  & 49.05\% & 29.82\% & 22.00\% \\
Dr.GRPO
  & Token-level PPO clip ($\epsilon$=0.2), length-unbiased reduction, no std norm
  & 52.42\% & 33.58\% & 25.13\% \\
GRPO
  & Token-level PPO clip ($\epsilon$=0.2), seq-mean reduction, std norm
  & 47.59\% & 30.89\% & 22.20\% \\
\midrule
\multicolumn{5}{l}{\textit{Single-factor ablation from GSPO}} \\
\midrule
Length-unbiased reduction
  & \texttt{loss\_reduction}: seq\_mean $\to$ seq\_mean\_token\_sum\_norm
  & 42.02\% & 24.88\% & 18.53\% \\
Std normalization
  & \texttt{grpo\_norm\_by\_std}: false $\to$ true
  & 52.73\% & \textbf{34.37\%} & \textbf{26.41\%} \\
\bottomrule
\end{tabular}%
}
\end{table}